\documentclass{article}
\usepackage{float}
\usepackage{authblk}
\usepackage{algorithm}
\usepackage{algorithmicx}  
\usepackage{algpseudocode}  
\usepackage{setspace}
\usepackage{bm}

\usepackage{url}
\usepackage{amsmath}
\usepackage{stfloats}
\usepackage{color}
\usepackage{mathrsfs}
\usepackage{dcolumn}
\usepackage{subfigure}  
\usepackage{multirow}
\usepackage{graphics}
\usepackage{graphicx}
\usepackage{geometry}
\usepackage{amsfonts} 
\usepackage{amssymb}
\usepackage{amsthm}
\usepackage{booktabs}
\usepackage{longtable}

\definecolor{gry}{rgb}{0.6,0.6,0.6} 
\definecolor{red}{rgb}{1,0,0} 
\definecolor{mygreen}{rgb}{0,0.7,0} 
\usepackage[colorlinks,linkcolor=mygreen,anchorcolor=blue,citecolor=blue]{hyperref}
\title{\textbf{Computation of Reachable Sets Based on Hamilton-Jacobi-Bellman Equation with Running Cost Function}\\
\normalsize{\textcolor{red}{This work has been submitted to the IEEE for possible publication. Copyright may be transferred without notice, after which this version may no longer be accessible.}}}

\author[a]{Weiwei Liao}
\author[b]{Tao Liang\thanks{Corresponding author: liang\_yit@outlook.com}}

\affil[a]{\footnotesize{College of Automation Engineering, 
Nanchang University, Nanchang, Jiangxi, China }}
\affil[b]{\footnotesize{
Yancheng Institute Of Technology, Yancheng, Jiangsu, China}}

\newtheorem{assumption}{\textbf{Assumption}}
\newtheorem{remark}{\textbf{Remark}}
\newtheorem{thm}{\textbf{Theorem}}
\newtheorem*{prf}{\textbf{Proof}}
\newtheorem{definition}{\textbf{Definition}}
\begin{document}
\maketitle
\begin{abstract}
	A novel method for computing reachable sets is proposed in this paper. 
	In the proposed method, a Hamilton-Jacobi-Bellman equation with running cost function 
	is numerically solved and the reachable sets of different time horizons are characterized  
	by a family of non-zero level sets of the solution of the Hamilton-Jacobi-Bellman equation.
	In addition to the classical reachable set, by setting different running cost functions and terminal conditions 
	of the Hamilton-Jacobi-Bellman equation, the proposed method 
	allows to compute more generalized reachable sets, which are referred to as cost-limited reachable sets.
	In order to overcome the difficulty of solving the Hamilton-Jacobi-Bellman equation caused by 
	the discontinuity of the solution, a method based on recursion and grid interpolation is employed.
	At the end of this paper, some examples are taken
	to illustrate the validity and generality of the proposed method.
\end{abstract}
\quad \\
\textbf{Keywards:} Reachability, Hamilton-Jacobi-Bellman equation, Running cost function, Level set, Numerical method.

\section{Introduction}
The methods to analyze linear systems have been reasonably mature, some 
crucial characteristics such as stability, controllability and observability have been systematically 
and strictly defined. Also, the analyses on these characteristics have become the common steps in solving
many engineering problems. However, for nonlinear systems, these characteristics are quite difficult to
be defined and analyzed \cite{b1.1}, such that for some nonlinear systems, their behaviors are difficult 
to predict.

Reachability analysis is an effective method to study the behavior of nonlinear control systems. 
By applying reachability analysis, one can solve a variety of engineering problems, 
especially those involving system safety \cite{a1,a2,i8}, feasibility \cite{tcyb2}, and control law design \cite{a3,a4}. 
In reachability analysis, one specifies a set in the state space as the target set
and then aims to find a set of initial states of the trajectories that can reach the target set
within a given time horizon \cite{a5,a5.5}. Such a set is referred to as the reachable set. 
However, finding reachable sets is a challenging task, which involves various aspects such as computation and data storage.
The most intuitive approach is to verify each point in the state space one by one, however, 
this approach often consumes a lot of time due to the diversity of system states and control inputs \cite{i1,a6}.
Therefore, formal verification methods are needed.

Unfortunately, it is nearly impossible to compute reachable sets using a analytical approach, except for some special problems \cite{a7}.
In recent years, various numerical methods have been proposed, which can be divided into two categories: 
the Lagrangian methods \cite{i2,tcyb1,a8,isa1} and the methods based on state space discretization \cite{a8.5,a9,a11,i1,a1,a2}.
The former can solve the reachability problems of high-dimensional systems, 
but has high requirements on the form of the control system, and thus is mainly used to solve linear problems.
The latter has less requirements on the form of the control system and can therefore be used for nonlinear systems.
It is this universality that makes the methods based on state space discretization more widely used in engineering,
and the level set method \cite{a5,i1,a1,a2} is a representative of them.

In the level set method, a Hamilton-Jacobi-Bellman (HJB) equation \textit{without} running cost function 
is constructed, and the terminal condition of this equation is set to the signed distance function of the target set.
Then the state space is discretized into a Cartesian grid structure and the HJB equation is numerically solved. 
During the computation, the values of the equation's solutions at the grid points are stored in 
an array that has the same dimensions as the state space. Finally, the reachable set is characterized as the zero-level set of the solution.

The principle of the level set method determines that its storage space requirements are quite demanding.
To save the reachable set of a given time horizon, one needs to save the solution of the HJB equation at a certain time point,
the memory required grow significantly with an increase in the problem's dimension \cite{a3,a12,i5}. 
To save the reachable sets under different time horizons, one needs to save the solutions of the HJB equation 
at different time points, which in turn leads to a multiple increase in storage space consumption.
In addition, in the level set approach, saving the solutions of the HJB equation at multiple moments is 
also necessary for designing the control law \cite{a3,a5,a13}. These limitations restrict the development 
and application of the level set method to some extent.

In order to overcome the above-mentioned limitations, this paper proposes a new method to compute reachable sets.
In the proposed method, a HJB equation \textit{with} running cost function is numerically solved, and the 
reachable sets under different time horizons can be characterized by different non-zero level sets of 
the solution of the HJB equation at a certain time point. 
Such a mechanism can significantly reduce the consumption of storage space and facilitate the design of control law.
In addition, more generalized reachability problems can be solved by 
setting different running cost functions and terminal conditions. 
In these problems, a performance index can be constructed, which is a combination of
a Lagrangian (the time integral of a running cost) and an endpoint cost. 
The aim is to find a set of initial states of the trajectories that can reach the target set
before the performance index increasing to the given admissible cost.
In this paper, such a set is referred to as a cost-limited reachable set.

In summary, the main contributions of this paper are as follows:
\begin{itemize}
	\item [(1)] A novel method for computing reachable sets based on the HJB equation with a running cost function is proposed.
	This method can significantly reduce the storage space consumption and bring convenience to the control law design.
	\item [(2)] The reachability problem is generalized by setting different running cost functions and terminal conditions, 
	and the definition of cost-limited reachable set is put forward.
	\item [(3)]  To overcome the discontinuity of the solution of the HJB equation, 
	a numerical method based on recursion and grid interpolation is applied to solve 
	the HJB equation.
\end{itemize}

The structure of this paper is as follows. 
Section II briefly introduces the reachability problem and the level set method. 
Section III describes the method to construct the HJB equation of the proposed method 
and the representation of the reachable set. 
Section IV generalizes the reachability problem and 
presents the definition of cost-limited reachable set, and also introduces the method to design control law.
A method to solve the HJB equation is proposed in Section V and some
numerical examples are given in Section VI. The results are summarized
in Section VII.

\section{Preliminaries}
\subsection{Reachability problem}
Consider a continuous time control system with fully observable state:
\begin{align}
	\label{sys1}
	\dot{s}=f(s,u)
\end{align}
where $s\in\mathbb{R}^n$ is the system state, $u\in\mathcal{U}$ is referred to as the control input.
The function $f(.,.):\mathbb{R}^n\times \mathcal{U} \to \mathbb{R}^n$ is bounded. Let $\mathscr{U}$ denote the set of
lebesgue measurable functions from the time interval $[0,\infty)$ to $\mathcal{U}$. 
Then, given the initial state $s_{t_0}$ at time $t_0$, $u(.)\in\mathscr{U}$, the evolution of system (\ref{sys1})
in time interval $[t_0,t_1]$ can be denoted as a continuous trajectory $\phi_{t_0}^{t_1}(.,s_{t_0},u(.)):[t_0,t_1]\to \mathbb{R}^n$
and $\phi_{t_0}^{t_1}(t_0,s_{t_0},u(.))=s_{t_0}$. Given a target set $K$ and a time horizon $T$, 
The definition of reachable set is \cite{tcyb1,a5}:

\begin{definition}[Reachable set]
	\label{def1}
	\begin{align}
		\begin{split}
			\label{defreach}
		\mathcal{R}(K,T)=\left\{s_0\in\mathbb{R}^n| \exists t\in [0,{T}],\exists u(.)\in \mathscr{U},\right. \\
		\left. \phi_{0}^{T}(t,s_0,u(.))\in K  \right\}
		\end{split}
	\end{align}
\end{definition}

\subsection{Level set method}
In level set method, the following HJB equation about $V(.,.):\mathbb{R}^n \times \mathbb{R}$ is numerically solved:
\begin{align}
	\label{hjb1}
	\left\{ \begin{array}{l}
		\displaystyle{ \frac{\partial V}{\partial t}(s,t)+ 
		\min \left[0, \min_{u\in \mathcal{U}} \frac{\partial V}{ \partial s}(s,t)f(s,u)   \right]=0}\\
		\text{s.t.} V(s,T)=l(s)
	\end{array}\right.
\end{align}
where $l(.)$ is bounded and Lipschitz continuous, and satisfies $K=\left\{s\in\mathbb{R}^n|l(s) \leq 0 \right\}$.
The solution is approximated on a Cartesian grid of the state space.
The reachable set is represented as the zero level set of function $V(.,0)$, i.e.
\begin{align}
	\mathcal{R}(K,T)=\left\{s\in\mathbb{R}^n|V(s,0) \leq 0 \right\}
\end{align}
Based on this principle, several mature toolboxes have been developed \cite{tr1,a14,a2} and applied to many practical engineering problems, 
such as flight control systems \cite{a15,i6,i7,a4,a18}, 
ground traffic management systems \cite{a16,a17,i12}, air traffic management systems \cite{a1,a2,isa2}, etc.

\begin{figure}[H]
	\centering
	\includegraphics[width=0.42\textwidth]{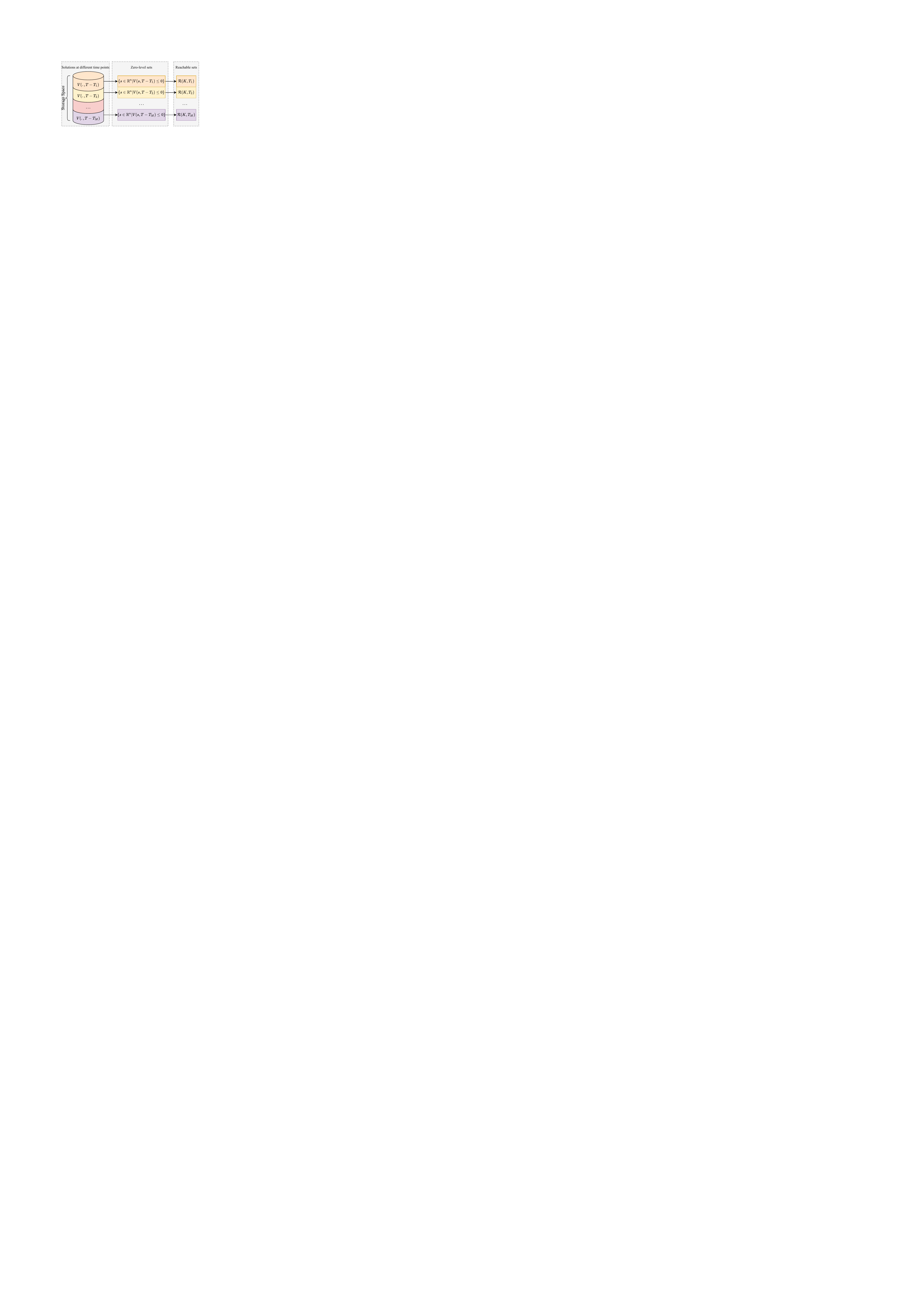}
	\caption{Method of saving reachable sets by the level set method.}
	\label{fig1}
\end{figure}

Denote the number of grid points in the $i$th dimension of the Cartesian grid as $N_i$, then the storage space consumed to save $\mathcal{R}(K,T)$ 
is proportional to $ \prod_{i=1}^n N_i $.
It should be noted that, for $T_1,...,T_M\in [0,\infty)$, 
the expressions of the reachable sets of these time horizons are as follows:
\begin{align}
	\begin{split}
		&\mathcal{R}(K,T_1)=\left\{s\in\mathbb{R}^n|V(s,T-T_1) \leq 0\right\}\\
		&...\\
		&\mathcal{R}(K,T_M)=\left\{s\in\mathbb{R}^n|V(s,T-T_M) \leq T_M\right\}
	\end{split}
\end{align}
Since the value functions $V(.,T-T_1),...,V(.,T-T_M)$ are each different,
the storage space consumption required to save these reachable
sets is proportional to $ M\prod_{i=1}^n N_i $, see Fig. \ref{fig1}. 

\section{Method Based on HJB Equation with Running Cost Function}
\subsection{Reachability problem and optimal control}
Consider a case where the control input $u(.)$ aims
to transfer the system state to the target set in the shortest possible time.
In this case, if a trajectory can reach the target set in the time horizon $T$, 
then the initial state of this trajectory belongs to the reachable set.
Therefore, a value function $\widetilde{W}(.):\mathbb{R}^n \to \mathbb{R}$ can be constructed as follows:


\begin{align}
	\label{minvalue}
	\widetilde{W}(s_0)=\begin{cases}
		\left\{\begin{array}{rl}
			\displaystyle{\min_{u(.)}}& \displaystyle{ \int_0^{t_f} \mathbb{I}(s(t)) dt } \\
			\text{s.t.}& \dot{s}(t)=f(s(t),u(t))\ \forall t\in [0,t_f]\\
			 \ &s(0)=s_0\\
			 \ &u(t)\in\mathcal{U} \ \forall t\in [0,t_f]\\
			 \ &s(t_f)\in K
		\end{array}\right\},\\ \ \ \ \ \   \text{if } \exists u(.)\in \mathscr{U}\ \exists t\in [0,\infty)\ 
		 \phi_{0}^{\infty}(t,s_0,u(.))\in K\\
	\infty,  \text{otherwise}
	\end{cases}
\end{align}
where $\mathbb{I}(.):\mathbb{R}^n \to \mathbb{R}$ is a running cost function and $\mathbb{I}(s)\equiv 1$.
then the reachable set $\mathcal{R}(K,T)$ can be characterized by the $T$-level set 
of $\widetilde{W}(.)$, i.e.
\begin{align}
	\mathcal{R}(K,T)=\left\{s\in\mathbb{R}^n|\widetilde{W}(s) \leq T\right\}
\end{align}


Define a modified dynamic system:
\begin{align}
	\label{mdfdyn}
	f_K(s,u)=\begin{cases}
		f(s,u),&s\notin K\\
		\mathbf{0},& s\in K
	\end{cases}
\end{align}
and a modified running cost function:
\begin{align}
	\label{mdfruc}
	\mathbb{I}_K(s)=\begin{cases}
		1,&s\notin K\\
		0,& s\in K
	\end{cases}
\end{align}
Given the state $s_{t_0}$ at time $t_0$, 
$u(.)\in\mathscr{U}$, the evolution of system (\ref{mdfdyn})
in time interval $[t_0,t_1]$ can be denoted as $\hat{\phi}_{t_0}^{t_1}(.,s_{t_0},u(.)):[t_0,t_1]\to \mathbb{R}^n$
and $\hat{\phi}_{t_0}^{t_1}(t_0,s_{t_0},u(.))=s_{t_0}$.

Based on system (\ref{mdfdyn}) and running cost (\ref{mdfruc}), we can also construct another value function:
\begin{align}
	\label{minvaluemd}
	\overline{W}(s_0,\bar{T})=
		\left\{\begin{array}{rl}
			\displaystyle{\min_{u(.)}}&  \displaystyle{\int_0^{\bar{T}} \mathbb{I}_K(s(t)) dt} \\
			\text{s.t.}& \dot{s}(t)=f_K(s(t),u(t))\ \forall t\in [0,\bar{T}]\\
			 \ &s(0)=s_0\\
			 \ &u(t)\in\mathcal{U} \ \forall t\in [0,\bar{T}]
		\end{array}\right.
\end{align}

Consequently, Eq. (\ref{minvalue}) and Eq. (\ref{minvaluemd}) have the following equivalence:
\begin{thm}\label{thm1}
	For any $s_0\in \left\{s\in\mathbb{R}^n | \overline{W}(s,\bar{T}) < \bar{T} \right\}$, 
	$\widetilde{W}(s_0)=\overline{W}(s_0,\bar{T})$ holds.
\end{thm}
\begin{prf}
	The maximum of the value function $\overline{W}(.,\bar{T})$ is:
	\begin{align}
		\max_{s_0\in\mathbb{R}^n} \overline{W}(s_0,\bar{T}) = \int_0^{\bar{T}} \max_{s\in\mathbb{R}^n} \mathbb{I}_K (s) dt = \bar{T}
	\end{align}
	The trajectories of system (\ref{mdfdyn}) correspond to the same trajectories as the evolution of system (\ref{sys1})
	as long as it evolves outside the target set, 
	once a trajectory of system (\ref{mdfdyn}) touches the border of the target set, it stays at the border of the target set and 
	the modifies running cost function is set to $0$. 
	Consequently, 
	\begin{align}
		&\overline{W}(s_0,\bar{T})<\bar{T} \Longleftrightarrow \exists u(.) \in \mathscr{U}\ \exists t_f\in [0,\bar{T}) \nonumber \\
		&\forall t\in [t_f,\bar{T}]\ \hat{\phi}_{0}^{\bar{T}}(t,s_{t_0},u(.)) \in K 
	\end{align}
	Therefore, 
	\begin{align}
		\begin{split}
			&\overline{W}(s_0,\bar{T})<\bar{T} \Longleftrightarrow \\
			&\overline{W}(s_0,\bar{T})=\min_{u(.)\in \mathscr{U}} \int_0^{\bar{T}} \mathbb{I}_K(s(t)) dt \\
			&= \min_{u(.)\in\mathscr{U}} \left[ \int_0^{t_f} \mathbb{I}(s(t)) dt + \int_{t_f}^{ \bar{T} } 0 dt \right] \\ 
			&=\min_{u(.)\in\mathscr{U}} \left[ \int_0^{t_f} \mathbb{I}(s(t)) dt  \right] = \widetilde{W}(s_0)
		\end{split}
	\end{align}
	\rightline{$\square$}
\end{prf}

Theorem \ref{thm1} states that in the region $\left\{s\in\mathbb{R}^n | \overline{W}(s,\bar{T}) < \bar{T} \right\}$, 
$\widetilde{W}(.)$ and $\overline{W}(.,\bar{T})$ are equal, and for any $T \in [0,\bar{T})$, 
the reachable set $\mathcal{R}(K,T)$ can also be expressed as:
\begin{align}
	\mathcal{R}(K,T)=\left\{s\in\mathbb{R}^n|\overline{W}(s,\bar{T}) \leq T\right\}
\end{align}

In addition, for $T_1,...,T_M\in [0,\bar{T})$, the reachable sets $\mathcal{R}(K,T_1),...,\mathcal{R}(K,T_M)$ 
can be represented as different level sets of the value function $\overline{W}(.,\bar{T})$, 
and simply save $\overline{W}(.,\bar{T})$ to save all these reachable sets, i.e.
\begin{align}
	\begin{split}
		&\mathcal{R}(K,T_1)=\left\{s\in\mathbb{R}^n|\overline{W}(s,\bar{T}) \leq T_1\right\}\\
		&...\\
		&\mathcal{R}(K,T_M)=\left\{s\in\mathbb{R}^n|\overline{W}(s,\bar{T}) \leq T_M\right\}
	\end{split}
\end{align}

\subsection{Construction of HJB equation}
Based on system (\ref{mdfdyn}) and running cost function (\ref{mdfruc}), 
an HJB equation with running cost function can be constructed: 
\begin{align}
	\label{hjb2}
	\begin{cases}
		\displaystyle{ \frac{\partial W}{ \partial t }(s,t)= 
		\min_{u\in\mathcal{U}} \left[ \frac{\partial W}{ \partial s }(s,t) f_K(s,u) +\mathbb{I}_K(s)  \right]}\\
		\text{s.t.  }W(s,0)=0
	\end{cases}
\end{align}

\begin{thm}\label{thm2}
	The solution of Eq. (\ref{hjb2}) at time $\bar{T}$ and $\overline{W}(.,\bar{T})$ are equivalent, i.e., 
	for any $s \in \mathbb{R}^n$, 
	\begin{align}
		W(s,\bar{T})=\overline{W}(s,\bar{T})
	\end{align}
\end{thm}

\begin{prf}
	When $\bar{T}=0$, the following equation holds:
	\begin{align}
		\label{thm1eq1}
		\overline{W}(s,0)=\int_0^0 \mathbb{I}_K (s(t)) dt =0 
	\end{align}
	From the definition of $\overline{W}(.,\bar{T})$, this function is a cost-to-go function on
	the time interval $[0,\bar{T}]$. According to Bellman's principle of optimality \cite{a19}, 
	for any $t\in [0,\infty)$ a small enough $\Delta t$, the cost-to-go function should 
	satisfy the following equation:
	\begin{align}
		\label{thm1eq2}
		\begin{split}
			&\overline{W}(s(t),t+\Delta t)=\\ 
			&\min_{u\in\mathcal{U}} 
			\left[ \overline{W}(s(t+\Delta t),t)  + \int_t^{t+\Delta t} \mathbb{I}_K (s(\tau)) d\tau   \right]
		\end{split}
	\end{align}
	Since
	\begin{align}
		\label{thm1eq3}
		\begin{split}
			\overline{W}(s(t),t+\Delta t)=&\overline{W}(s(t),t)+\frac{\partial W }{\partial t} (s(t),t) \Delta t\\
			\overline{W}(s(t+\Delta t),t)=&\overline{W}(s(t),t)\\
			&+\frac{\partial W }{\partial s}(s(t),t) f_K(s(t),u(t)) \Delta t\\
			\int_t^{t+\Delta t} \mathbb{I}_K (s(\tau)) d\tau&=\mathbb{I}_K(s(t))\Delta t
		\end{split}
	\end{align}
	Substituting Eq. (\ref{thm1eq3}) into Eq. (\ref{thm1eq2}) yields the following equation:
	\begin{align}
		\label{thm1eq4}
		\frac{\partial \overline{W} }{ \partial t }(s,t)= 
		\min_{u\in\mathcal{U}} \left[ \frac{\partial \overline{W}}{ \partial s }(s,t) f_K(s,u) +\mathbb{I}_K(s)  \right]
	\end{align}
	Combining Eq. (\ref{thm1eq1}) and Eq. (\ref{thm1eq4}), the form is exactly the same as that of Eq. (\ref{hjb2}).

	\rightline{$\square$}
\end{prf}

\begin{figure}[H]
	\centering
	\includegraphics[width=0.42\textwidth]{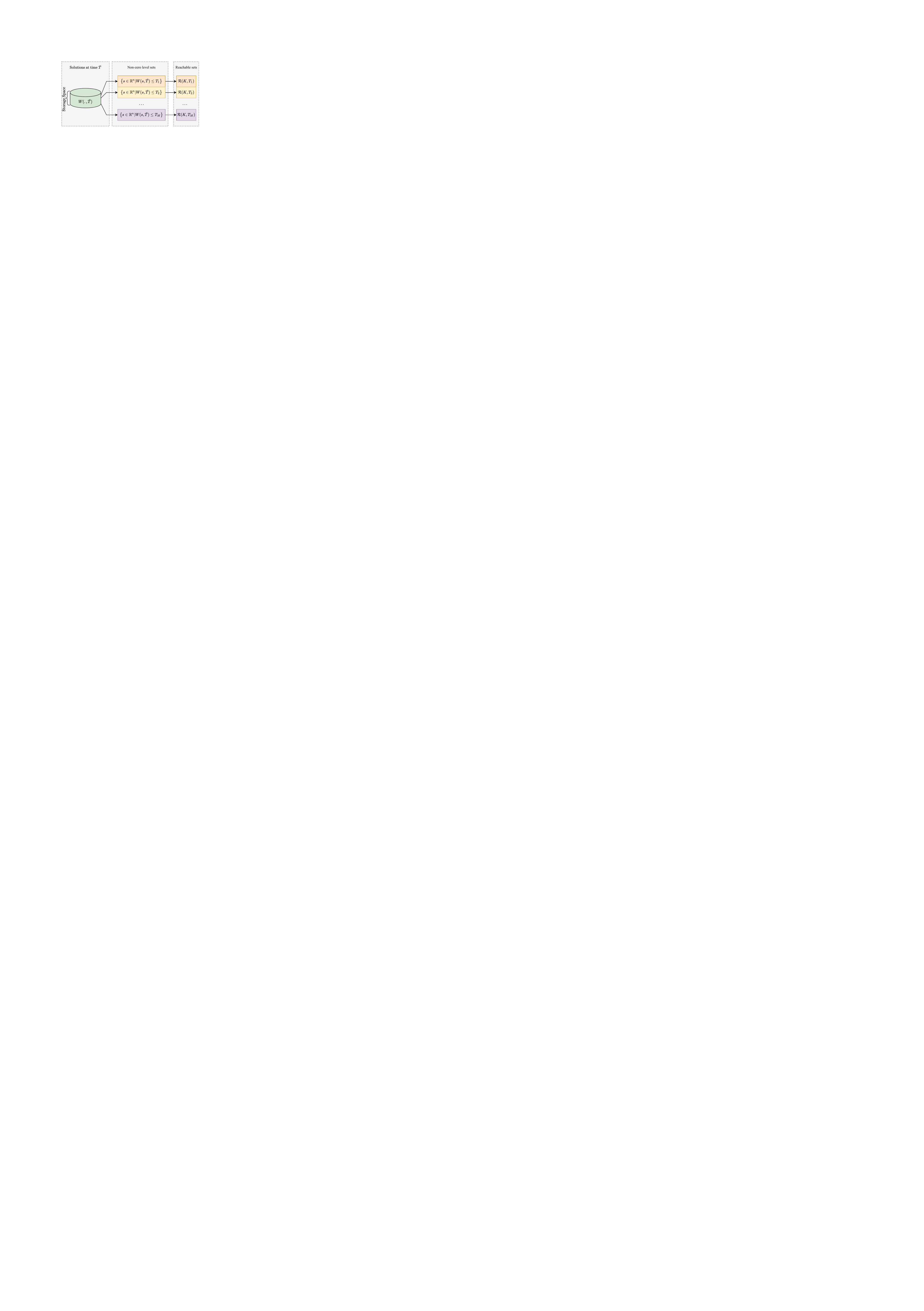}
	\caption{Method of saving reachable sets by the proposed method.}
	\label{fig2}
\end{figure}

Theorem \ref{thm2} indicates that the reachable sets of different time horizons
can be represented by different level sets of the solution of Eq. (\ref{hjb2}), i.e. 
\begin{align}
	\begin{split}
		&\mathcal{R}(K,T_1)=\left\{s\in\mathbb{R}^n|{W}(s,\bar{T}) \leq T_1\right\}\\
		&...\\
		&\mathcal{R}(K,T_M)=\left\{s\in\mathbb{R}^n|{W}(s,\bar{T}) \leq T_M\right\}
	\end{split}
\end{align}
Fig. \ref{fig2} illustrates the way to save the reachable sets by the proposed method.

\section{Generalization of reachability problems}
In the previous section, the running cost function and the terminal condition of HJB equation are quite special.
In fact, the HJB equation can also be 
used for more general reachability problems by setting different running cost functions 
and terminal conditions. This section introduces a novel type of reachability problems. 

\subsection{Definition of cost-limited reachable set}
A general running cost function is a scalar function of state and control input, denoted as
\begin{align}
    \label{runningcost2}
    c(.,.):\mathbb{R}^n \times \mathcal{U}\to \mathbb{R}
\end{align}

In this section, we assume that:
\begin{assumption}\label{ass1}
	$ \displaystyle{\min_{s\in\mathbb{R}^n,u\in\mathcal{U}}} c(s,u) = \lambda$ holds, where $\lambda$ is a positive real
	number.  
\end{assumption}
The above-mentioned assumption is easily satisfied in engineering practice,
such as the fuel consumption and path length per unit time are positive.

The performance index of the evolution of system (\ref{sys1}) initialized from $s_{t_0}$ at time $t_0$ under 
control input $u(.)$ in time interval $[t_0,t_1]$ is denoted as:
\begin{align}
	\begin{split}
	\mathcal{J}_{t_0}^{t_1}(s_{t_0},u(.))=\int_{t_0}^{t_1} c\left( \phi_{t_0}^{t_1}(t,s_{t_0},u(.)),u(t) \right) dt\\
	+\Phi\left( \phi_{t_0}^{t_1}(t_1,s_{t_0},u(.)) \right)
	\end{split}
\end{align} 
where $\Phi(.):\mathbb{R}^n\to \mathbb{R}$ is the endpoint cost function.
Given a target set $K$ and an admissible cost $J$, the cost-limited reachable set can be defined:
\begin{definition}[Cost-limited reachable set]
	\label{def2}
	\begin{align}
		\begin{split}
		\label{defcostreach}
		\mathcal{R}_c(K,J)=\left\{s_0\in\mathbb{R}^n| \exists t\in [0,\infty),\exists u(.)\in \mathscr{U},\right. \\
		\left. \phi_{0}^{\infty}(t,s_0,u(.))\in K \land \mathcal{J}_0^t(s_0,u(.)) \leq J \right\}
		\end{split}
	\end{align}
\end{definition}
where "$\land$" is the logical operator "AND".

Informally, under Assumption \ref{ass1}, the performance index increases with the increasing of time, 
the cost-limited reachable set is a set of initial states of trajectories that can be reach 
the target set before the performance index increasing to the given admissible cost.

\begin{figure}[H]
	\centering
	\includegraphics[width=0.3\textwidth]{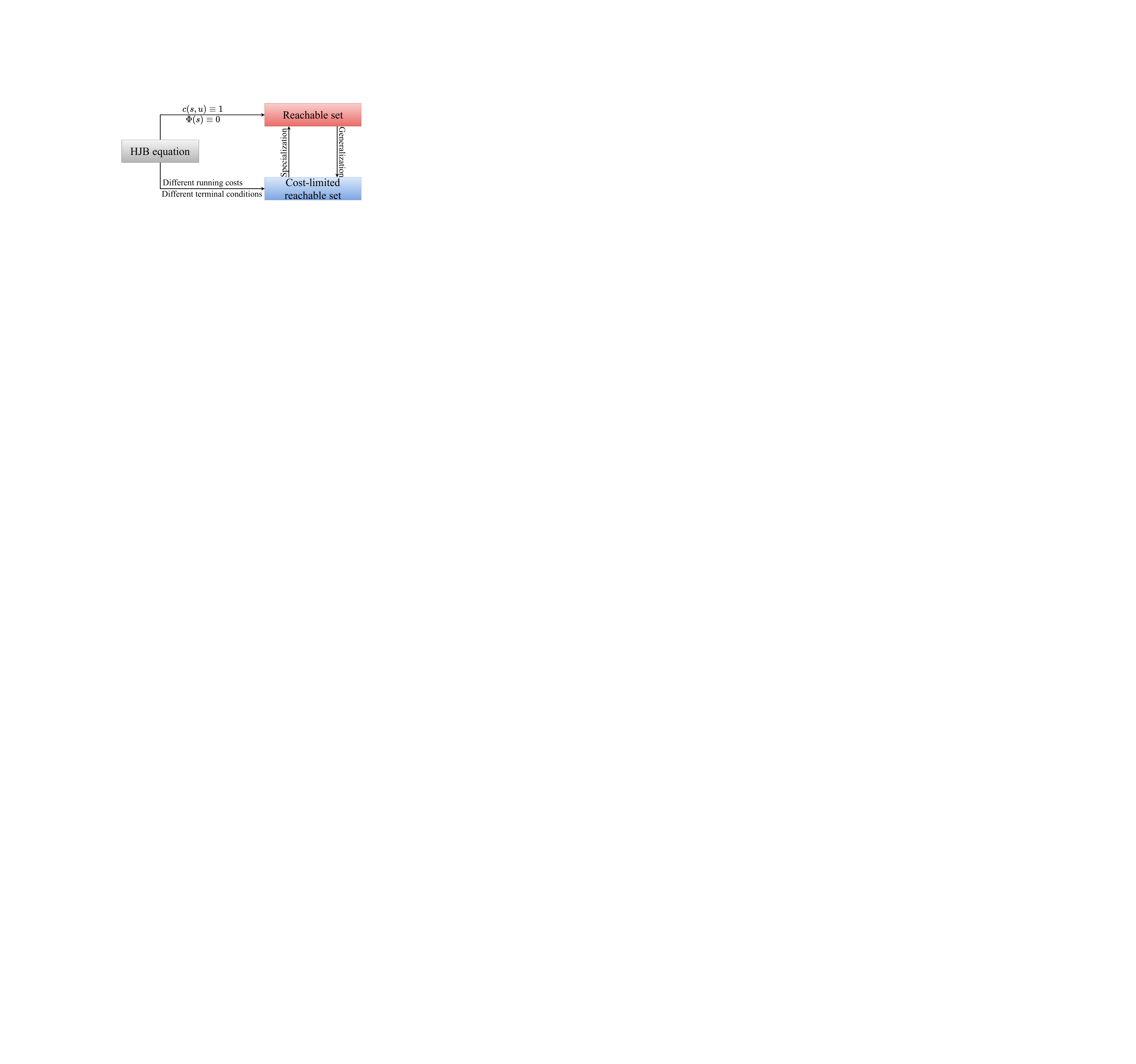}
	\caption{Connection between reachable set and cost-limited reachable set}
	\label{fig3}
\end{figure}

\begin{remark}
	According to Definition \ref{def1} and Definition \ref{def2}, reachable set is a special form of cost-limited reachable set. 
	If $c(s,u)\equiv 1$ and $ \Phi(s)\equiv 0$, then the performance index $ \mathcal{J}_0^t(s_0,u(.))=t$ and
	the cost-limited reachable set is degenerated into the reachable set, see Fig. \ref{fig3}. 
\end{remark}

\subsection{Computation of cost-limited reachable set}
Consider a case where the controller aims to transfer the system state to the target set with the least possible cost.
If a trajectory can enter the target set before the performance index increasing to the given admissible cost, 
its initial state must in the cost-limited reachable set. Similar to Eq. (\ref{minvalue}), 
a value function can be constructed as follows:
\begin{align}
	\label{minvalue2}
	\widetilde{W}_c(s_0)=\begin{cases}
		\left\{\begin{array}{rl}
			\displaystyle{\min_{u(.)}}& \left[\displaystyle{ \int_0^{t_f} c(s(t),u(t)) dt+ \Phi(s(t_f)) }  \right]   \\
			\text{s.t.}& \dot{s}(t)=f(s(t),u(t))\ \forall t\in [0,t_f]\\
			 \ &s(0)=s_0\\
			 \ &u(t)\in\mathcal{U} \ \forall t\in [0,t_f]\\
			 \ &s(t_f)\in K
		\end{array}\right\},\\ \ \ \ \   \text{if } \exists u(.)\in \mathscr{U}\ \exists t\in [0,\infty)\ 
		\phi_{0}^{\infty}(t,s_0,u(.))\in K\\
   \infty,  \text{otherwise}
	\end{cases}
\end{align}
The cost-limited reachable set $\mathcal{R}_c(K,J)$ can be represented as the $J$-level set of function $\widetilde{W}_c(.)$, i.e.
\begin{align}
	\mathcal{R}_c(K,J)=\left\{s\in\mathbb{R}^n|\widetilde{W}_c(s)\leq J \right\}
\end{align}


Construct a modified running cost function on the basis of Eq. (\ref{runningcost2}):
\begin{align}
    \label{mdfrc2}
    c_K(s,u)=\begin{cases}
        c(s,u),&s\notin K\\
        0, & s\in K
    \end{cases}
\end{align}
Similar to Eq. (\ref{minvaluemd}), based on the modified running cost function (\ref{mdfrc2}) 
and the modified system (\ref{mdfdyn}), the following value function can be constructed:
\begin{align}
	\label{minvaluemd2}
	\overline{W}_c(s_0,\bar{T})=
		\left\{\begin{array}{rl}
			\displaystyle{\min_{u(.)}}&  \left[\displaystyle{ \int_0^{\bar{T}} c_K(s(t),u(t)) dt+ \Phi(s(\bar{T})) }  \right] \\
			\text{s.t.}& \dot{s}(t)=f_K(s(t),u(t))\ \forall t\in [0,\bar{T}]\\
			 \ &s(0)=s_0\\
			 \ &u(t)\in\mathcal{U} \ \forall t\in [0,\bar{T}]
		\end{array}\right.
\end{align}

Denote $\min_{s\in\mathbb{R}^n}\Phi(s)=\Lambda$, then Eq. (\ref{minvalue2}) and Eq. (\ref{minvaluemd2})
have the following equivalence: 
\begin{thm}\label{thm3}
	For any $s_0\in \left\{s \in\mathbb{R}^n| \overline{W}_c(s_0,\bar{T})< \lambda \bar{T} + \Lambda  \right\}$, 
	$\widetilde{W}_c(s_0)=\overline{W}_c(s_0,\bar{T})$ holds.
\end{thm}

\begin{prf}
	The modified running cost function (\ref{mdfrc2}) is not less than $\lambda$ when the trajectory of system (\ref{mdfdyn}) evolves 
    outside the target set $K$. Therefore, 
    \begin{align}
        \begin{split}
            &\forall u(.)\ \forall t\in [0,\bar{T}]\ \hat{\phi}_{0}^{\bar{T}}(t,s_0,u(.))\notin K \Longrightarrow \\
            &\overline{W}_c(s_0,\bar{T}) \geq \int_0^{\bar{T}} \lambda dt + \Lambda = \lambda \bar{T}+\Lambda
        \end{split}
    \end{align}
    Consequently,
    \begin{align}
        \begin{split}
            &\overline{W}_c(s_0,\bar{T}) < \lambda \bar{T}+\Lambda \Longrightarrow\\
            &\exists u(.)\ \exists t_f\in [0,\bar{T})\ \forall t \in [t_f,\bar{T}]\ \hat{\phi}_{0}^{\bar{T}}(t,s_0,u(.))\in K
        \end{split}
    \end{align}
    Finally,
	\begin{align}
		\begin{split}
			&\overline{W}_c(s_0,\bar{T})<\lambda \bar{T}+\Lambda \Longrightarrow \\
			&\overline{W}_c(s_0,\bar{T})\\
            &=\min_{u(.)\in \mathscr{U}} \left[ \int_0^{\bar{T}} c_K(s(t),u(t)) dt +\Phi(s(\bar{T})) \right] \\
			&= \min_{u(.)\in\mathscr{U}} \left[ \int_0^{t_f} c(s(t),u(t)) dt + \int_{t_f}^{ \bar{T} } 0 dt + \Phi(s(\bar{T})) \right] \\ 
            &= \min_{u(.)\in\mathscr{U}} \left[ \int_0^{t_f} c(s(t),u(t)) dt + \int_{t_f}^{ \bar{T} } 0 dt + \Phi(s(\bar{t_f})) \right] \\ 
			&=\min_{u(.)\in\mathscr{U}} \left[ \int_0^{t_f} c(s(t),u(t)) dt + \Phi(s(\bar{t_f}))  \right] = \widetilde{W}_c(s_0)
		\end{split}
	\end{align}
    \rightline{$\square$}
\end{prf}

\begin{thm}\label{thm4}
	the value function $\overline{W}_c(.,\bar{T})$ can be 
	obtained by solving the following HJB equation about $W_c(.,.):\mathbb{R}^n\times \mathbb{R} \to \mathbb{R}$: 
	\begin{align}
		\label{hjb4}
		\begin{cases}
			\displaystyle{ \frac{\partial W_c}{ \partial t }(s,t)= 
			\min_{u\in\mathcal{U}} \left[ \frac{\partial W_c}{ \partial s }(s,t) f_K(s,u) +c_K(s,u)  \right]}\\
			\text{s.t.  }W_c(s,0)=\Phi(s_0)
		\end{cases}
	\end{align}
\end{thm}
The proof of Theorem \ref{thm4} is similar to that of Theorem \ref{thm2} and will not be repeated here.

\begin{figure}[H]
	\centering
	\includegraphics[width=0.42\textwidth]{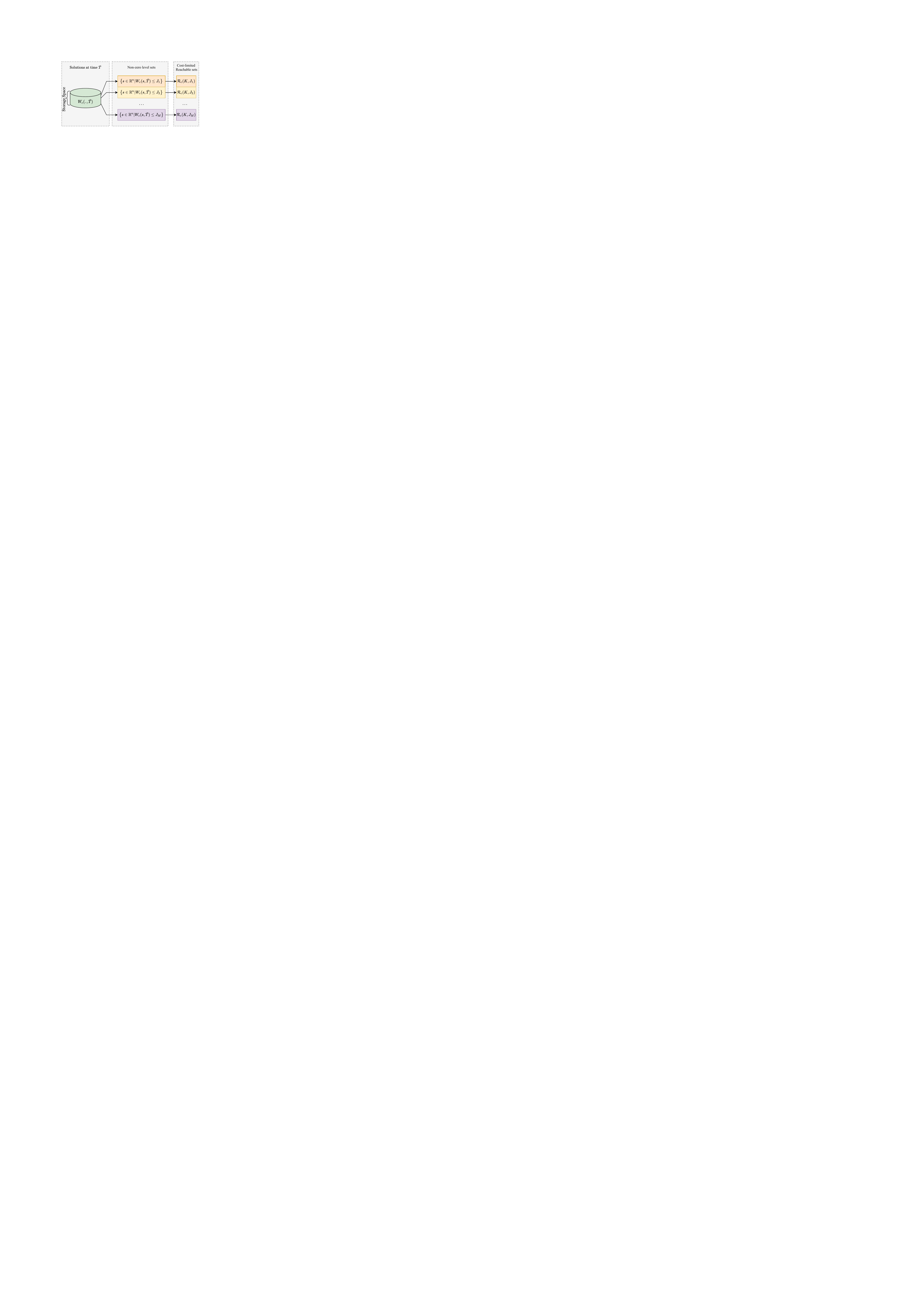}
	\caption{Method of saving cost-limited reachable sets by the proposed method.}
	\label{fig4}
\end{figure}

It follows from Theorem \ref{thm3} and Theorem \ref{thm4} that, for $J_1,...,J_M < \lambda \bar{T}+\Lambda $, 
the cost-limited reachable sets $\mathcal{R}_c(K,J_1),...,\mathcal{R}_c(K,J_M)$
can be represented as different level sets of $W_c(.,\bar{T})$ and all these sets can be saved by saving $W_c(.,\bar{T})$, i.e.
\begin{align}
    \begin{split}
        &\mathcal{R}_c(K,J_1)=\left\{s\in\mathbb{R}^n| W_c(s,\bar{T}) \leq J_1 \right\}\\
        &...\\
        &\mathcal{R}_c(K,J_M)=\left\{s\in\mathbb{R}^n| W_c(s,\bar{T}) \leq J_M \right\}
    \end{split}
\end{align}
See Fig. \ref{fig4}.

\subsection{Control law design}
\subsubsection{Control law in the level set method}
In the level set method, at time $t$, the optimal control input at state $s$ is \cite{a5,a3}:
\begin{align}
    \label{cl1}
    u^*(s,t)=\arg\min_{u\in\mathcal{U}} \frac{\partial V}{ \partial s }(s,t) f(s,u)
\end{align}
For any $s_0 \in \mathcal{R}(K,T)$, the trajectory initialized from $s_0$ can reach the target set $K$ 
under control law (\ref{cl1}) in time $T$.
Since $V(s,t)$ varies with time $t$, it is required to save $V(.,.)$ at each time point $t$ in time interval $[0,T]$ 
to implement this control law. This also requires a large amount of storage space.

\subsubsection{Control law in the proposed method}
The following control law ensures that the trajectory of the system enters the target set at the smallest performance index:
\begin{align}
    \label{cl2}
    u^*(s)=\arg\min_{u\in\mathcal{U}} \left[ \frac{\partial \widetilde{ W}_c}{ \partial s }(s) f(s,u)+c(s,u) \right] 
\end{align}
According to Theorem \ref{thm3} and Theorem \ref{thm4}, 
for any $s_0\in \left\{s\in\mathbb{R}^n| W_c(s,\bar{T}) < \lambda \bar{T}+\Lambda  \right\}$, 
$W_c(s_0,\bar{T})=\widetilde{W}_c(s_0)$. 

Consequently, 
for any $s_0\in \left\{s\in\mathbb{R}^n| W_c(s,\bar{T}) < \lambda \bar{T}+\Lambda  \right\}$, 
Eq. (\ref{cl2}) can be rewritten as:
\begin{align}
	\label{cl3}
    u^*(s_0)=\arg\min_{u\in\mathcal{U}} \left[ \frac{\partial  W_c}{ \partial s }(s_0,\bar{T}) f(s_0,u)+c(s_0,u) \right] 
\end{align}
All that needs to be saved to implement control law (\ref{cl3}) is the solution of HJB equation (\ref{hjb4}) at time $\bar{T}$.
Therefore, control law (\ref{cl3}) significantly reduces the storage space consumption compared to control law (\ref{cl1}).

\section{Method to Solve HJB Equation}

\begin{algorithm*}[tbp]
    \caption{Method to compute cost-limited reachable set}
	\label{alg1}
	\begin{algorithmic}[1]
		\State \textbf{Inputs:} Dynamic system (\ref{sys1}), set of achievable control input $\mathcal{U}$, 
		running cost function $c(.,.)$, endpoint cost function $\Phi(.)$,
		admissible costs $J_1,...,J_M$, target set $K$, number of time steps $m$, 
		computational domain $\Omega=[x_{\text{min}},x_{\text{max}}]\times [y_{\text{min}},y_{\text{max}}]$, 
		number of grids $N_x\times N_y$; 
		\State $\displaystyle{ \lambda\leftarrow \min_{s\in\mathbb{R}^n,u\in\mathcal{U}} c(s,u) }$, 
		$\displaystyle{ \Lambda \leftarrow \min_{s\in\mathbb{R}^n} \Phi(s) }$;

		\State $J_{\text{max}}\leftarrow\max\left( J_1,...,J_M \right)$, $\displaystyle{\bar{T}\leftarrow \frac{J_{\text{max}}-\Lambda}{\lambda}+\epsilon} $; 
		\textcolor{gry}{$\backslash\backslash$ $\epsilon$ 
		is a small positive number to ensure $\lambda\bar{T}+\Lambda > J_{\text{max}} $.}

		\State $\Delta t\leftarrow \displaystyle{ \frac{\bar{T}}{m} }$, 
		$\Delta x\leftarrow \displaystyle{ \frac{ x_{\text{max}}-x_{\text{min}} }{N_x-1} } $,
		$\Delta y\leftarrow \displaystyle{ \frac{ y_{\text{max}}-y_{\text{min}} }{N_x-1} } $;

		\State Construct the modified system in discretized form (\ref{dissys2}) and the modified system in discretized form (\ref{rcdis});

		\State Let $\mathcal{W}_c$ and $\mathcal{W}_c'$ be two $N_x\times N_y$ arrays;

		\For{$i\leftarrow 0,...,N_x-1$}\textcolor{gry}{\quad\quad\quad\quad $\backslash\backslash$ 
		Set the terminal condition of the HJB equation to the endpoint cost function.}
			\For{$j\leftarrow 0,...,N_y-1$}
				\State $s_0\leftarrow [x_{\text{min}}+i\Delta x,y_{\text{min}}+j\Delta y ]^\mathrm{T}$;
				\State $\mathcal{W}_c[i][j]\leftarrow \Phi(s_0)$;
			\EndFor
		\EndFor

		\For{$k\leftarrow 1,...,m$}
			\State Construct a bilinear interpolation function $\widehat{W}_c(.)$ using $\mathcal{W}$;
			\For{$i\leftarrow 0,...,N_x-1$}
				\For{$j\leftarrow 0,...,N_y-1$}
					\State $s_0\leftarrow [x_{\text{min}}+i\Delta x,y_{\text{min}}+j\Delta y ]^\mathrm{T}$;
					\State $\mathcal{W}_c'[i][j]\leftarrow \displaystyle{ \min_{u\in\mathcal{U}} \left[  C_K(s_0,u) + \widehat{W}_c(F_K(s_0,u))  \right] }$;
				\EndFor
			\EndFor
			\State Copy $\mathcal{W}_c'$ to $\mathcal{W}_c$;
		\EndFor
		\State Construct a bilinear interpolation function $\widehat{W}_c(.)$ using $\mathcal{W}$;
		\State $\mathcal{R}_c(K,J_1)\leftarrow \left\{s\in\mathbb{R}^n|\widehat{W}_c(s)\leq J_1 \right\},...,
		\mathcal{R}_c(K,J_M)\leftarrow \left\{s\in\mathbb{R}^n|\widehat{W}_c(s)\leq J_M \right\}$;
		\State \textbf{Return} $\mathcal{R}_c(K,J_1),..,\mathcal{R}_c(K,J_M)$;
    \end{algorithmic}
\end{algorithm*}

Since Eq. (\ref{hjb2}) is a special form of Eq. (\ref{hjb4}), this section introduces the method of solving Eq. (\ref{hjb4}).
The analytical solution of Eq. (\ref{hjb4}) is usually difficult to obtain. 
To make matters worse, the solution of Eq. (\ref{hjb4}) is not everywhere differentiable and sometimes even discontinuous, 
which leads to the difficulty in obtaining the viscosity solution as well \cite{i9}.
In the current research, a numerical method based on recursion and interpolation is introduced.

\subsection{Recursive formula of the solution}
Divide the time interval $[0,\bar{T}]$ into $m$ subintervals of length $\Delta t=\frac{\bar{T}}{m}$.
If $\Delta t$ is small enough, $u(.)$ can be regarded as a constant in interval $[k\Delta t,(k+1)\Delta t]$ 
for $k\in\mathbb{N}$, and system (\ref{sys1}) can be converted into the following discretized form:
\begin{align}
	s( (k+1)\Delta t )=F(s(k\Delta t),u(k\Delta t))
\end{align}
Thus, the discrete form of system (\ref{mdfdyn}) is:
\begin{align}
	\label{dissys2}
	\begin{split}
		s( (k+1)\Delta t )&=F_K(s(k\Delta t),u(k\Delta t))\\
		&=\begin{cases}
			F(s(k\Delta t),u(k\Delta t)), & s(k\Delta t) \notin K\\
			s( k\Delta t ), &s(k\Delta t) \in K
		\end{cases}
	\end{split}
\end{align}

The definite integral of the running cost function (\ref{runningcost2}) over the time interval $[k\Delta t,(k+1)\Delta t]$ is
denoted as:
\begin{align}
	\begin{split}
		C(s(k\Delta t),u(k\Delta t))=\int_{k\Delta t}^{(k+1)\Delta t} c(s(\tau),u(k\Delta t)) d\tau
	\end{split}
\end{align} 
and the integral of the modified running cost function (\ref{mdfrc2}) is denoted as:
\begin{align}
	\label{rcdis}
	\begin{split}
		&C_k(s(k\Delta t),u(k\Delta t))\\
		&=\begin{cases}
			C(s(k\Delta t),u(k\Delta t)), & s(k\Delta t)\notin K\\
			0,&s(k\Delta t)\in K
		\end{cases}
	\end{split}
\end{align} 

The recursive formula of the solution of Eq. (\ref{hjb4}) is:
\begin{align}
	\begin{split}
		&W_c(s,0)=\Phi(s)\\
		&W_c(s,(k+1)\Delta t)=\min_{u\in\mathcal{U}} [ C_k(s,u)+ \\ 
		& \ \ \ \ \ \ \ \ \ \ \ \ \ \ \ \ \ \ \ \ \ \ \ \ \ \ \ \ \ \  W_c(F_k(s,u),k\Delta t )   ]
	\end{split}
\end{align}

\subsection{Approximation of the solution}

This subsection introduces a method based on interpolation to approximate $W_c(.,k\Delta t) $.
The proposed method is similar to the level set method in this respect. 
First, a rectangular computational domain, denoted as $\Omega$, needs to be specified in the state space and
divided into a Cartesian grid structure. 
The value of the solution at the grid point is stored in an array with the same dimensions as the state space,
and $W_c(.,k\Delta t) $ is approximated by the grid interpolation.
Take a two-dimensional system as an example, and denote the system state as $s=[x,y]^\mathrm{T}$.
The pseudocode of the proposed method is shown in Algorithm \ref{alg1}.

\section{Numerical Examples}
This section provides two examples, the first one about the reachable set of a two-dimensional system, 
to visually demonstrate the superiority of the proposed method in terms of storage space consumption.
The second example is about the cost-limited reachable sets in a practical engineering problem 
to demonstrate the generality of the proposed method.

\subsection{Two-dimensional system example}
\begin{figure}[H]
	\centering
	\subfigure[$\mathcal{R}(K,T_1)$]{\includegraphics[width=0.24\textwidth]{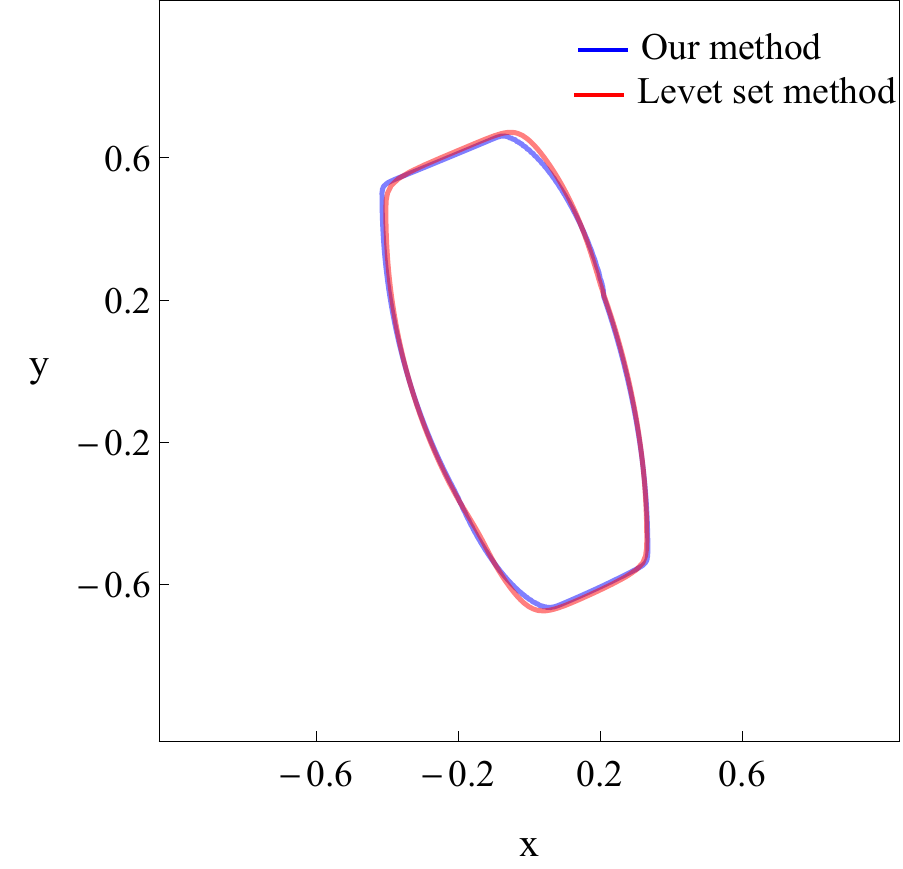}}
	\subfigure[$\mathcal{R}(K,T_2)$]{\includegraphics[width=0.24\textwidth]{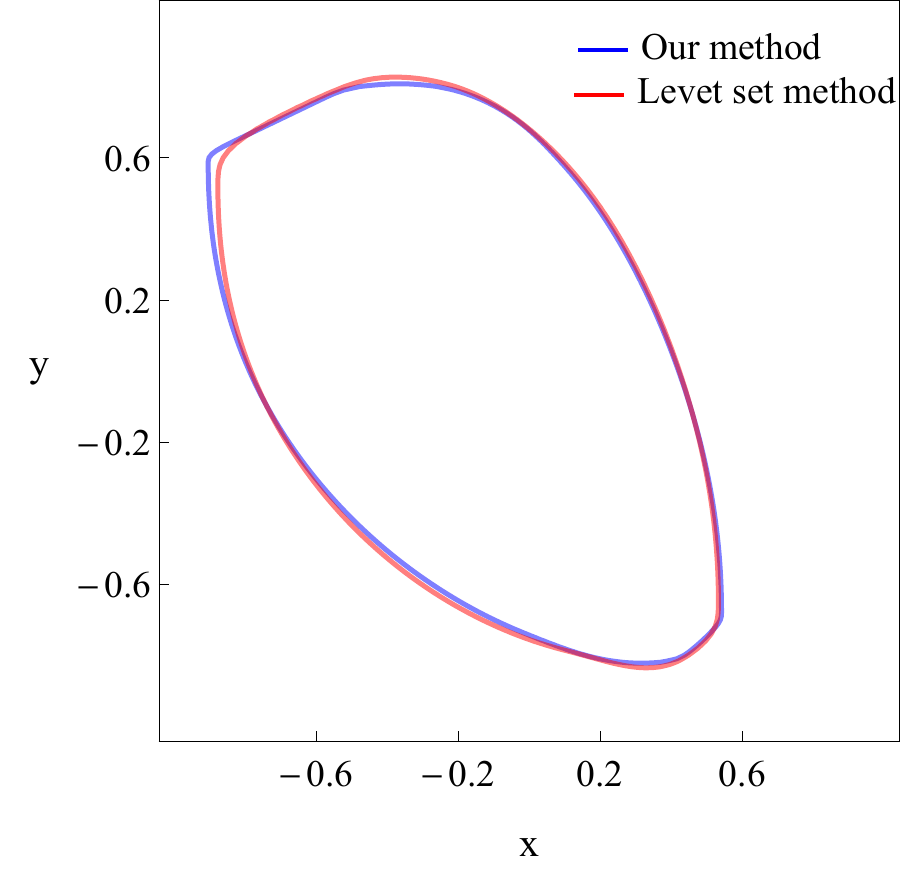}}\\
	\subfigure[$\mathcal{R}(K,T_3)$]{\includegraphics[width=0.24\textwidth]{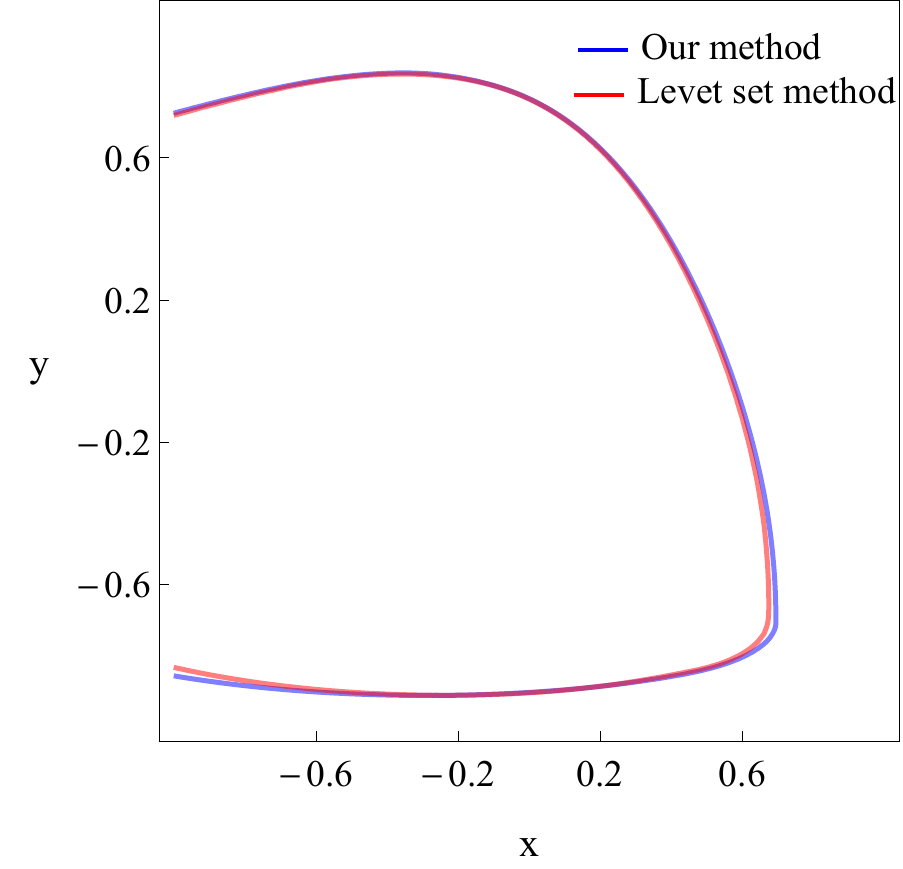}}
	\subfigure[$\mathcal{R}(K,T_4)$]{\includegraphics[width=0.24\textwidth]{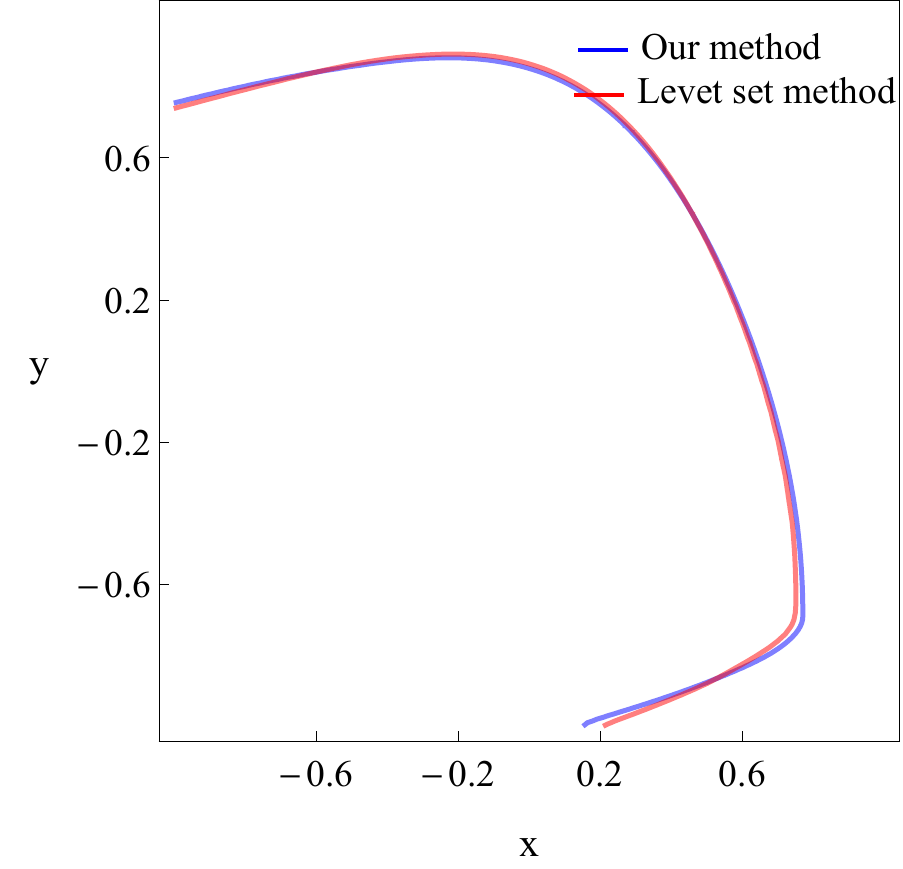}}
	\caption{Computation results of the reachable sets.}
	\label{fig5}
\end{figure}

Consider the following system:
\begin{align}
	\dot{s}=\left[
		\begin{array}{c}
			\dot{x}\\ \dot{y}
		\end{array}
	\right]=\left[
		\begin{array}{c}
			y+x^2 \\ -x+y^3+u
		\end{array}
	\right]=f(s,u)
\end{align}
where $s=[x,y]^\mathrm{T}$ is the system state, $u\in\mathcal{U}=[-1,1]$ is the control input. 
The target set $K=[-0.2,0.2]\times [-0.2,0.2]$. The given time horizons are $T_1=0.5,T_2=1,T_3=1.5,T_4=2$.
The task is to compute the reachable set corresponding to each time horizon. 
Table \ref{tb1} outlines the parameters that are specified for the reachable set
computation. 
\begin{table}[H]
	\centering
	\caption{Solver settings for the two-dimensional system example}
	\label{tb1}
	\begin{tabular}{lc}
	\toprule[1pt] 
	\textbf{Parameter} 			  			& \textbf{Setting}  \\ \toprule
	Computational domain $\Omega$         	&      $[-1,1]\times [-1,1]$      \\   
	Number of grid points $N_x\times N_y$   &      $201\times 201$     \\  
	$\bar{T}$               				&      $2.1$      \\    
	Number of time steps $m$            	&      $105$     \\ 
	Time step size $\Delta t$            	&      $0.02$     \\ \bottomrule[1pt] 
	\end{tabular}
\end{table}
As this example computes the reachable sets, the running cost function is set to $c(s,u)\equiv 1$ and 
the endpoint cost function is set to $\Phi(s)\equiv 0$. 
Fig. \ref{fig5} shows the computation results of the proposed method and compares them with those of the level set method
(The computational domain and the number of grids used in the level set method are the same as in our method,
and the terminal condition of the HJB equation in the level set method is set as $V(s,T_4)=l(s)$).

As can be seen in Fig. \ref{fig5}, the results of the proposed method and those of the level set method almost coincide,
which indicates that the proposed method has a high accuracy.

\begin{figure*}
	\centering
	\ \ \subfigure[The way the proposed method saves the reachable sets.]{\includegraphics[width=0.74\textwidth]{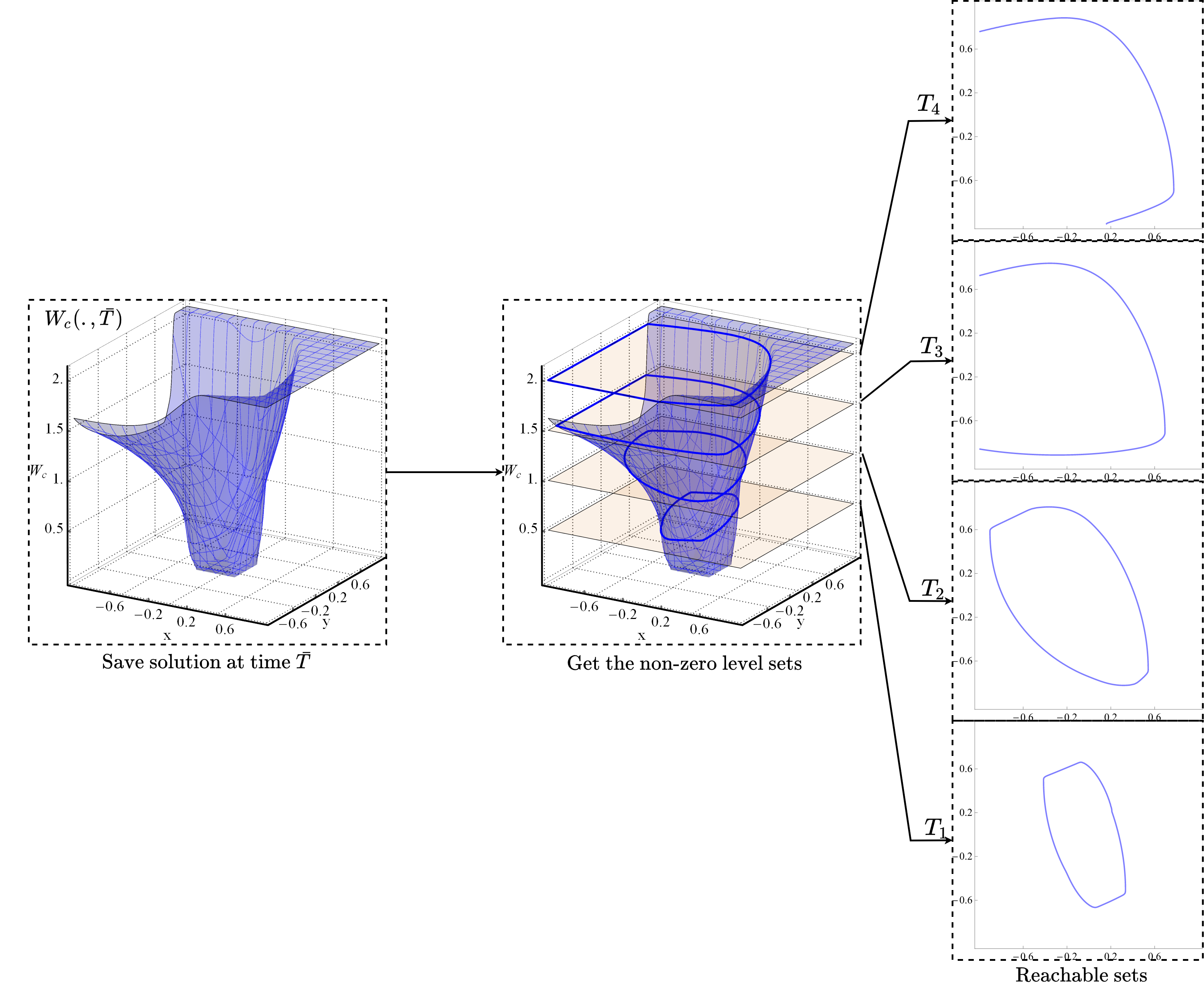}}\\
	\subfigure[The way the level set method saves the reachable sets.]{\includegraphics[width=0.75\textwidth]{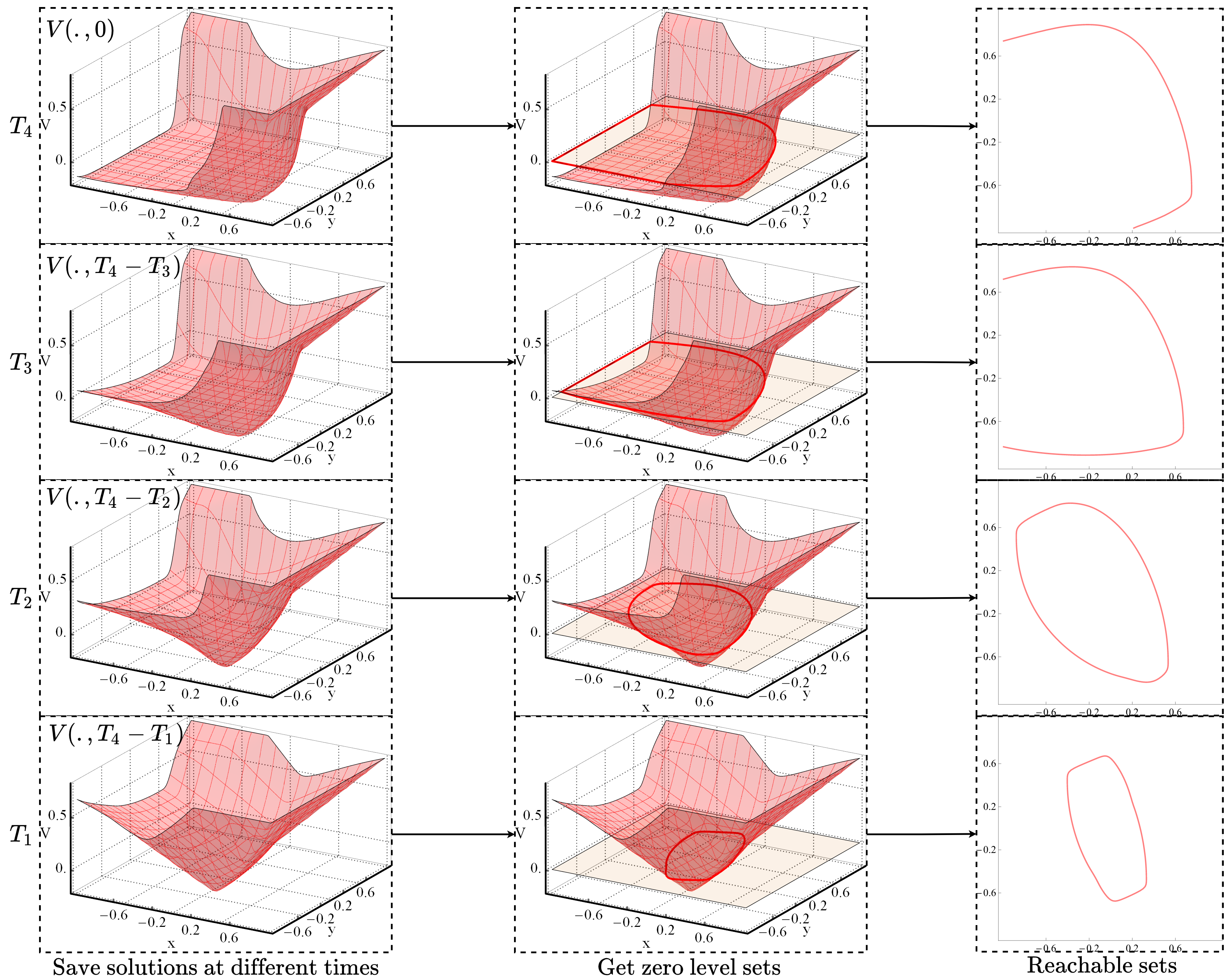}}
	\caption{Ways to save the reachable sets in different methods.}
	\label{fig6}
\end{figure*}

Fig. \ref{fig6} visualizes the storage forms of reachable sets in the proposed method as well as in the level set method. 
Our method only needs to save function $W_c(.,\bar{T})$, 
while the level set method needs to save $V(.,0)$, $V(.,T_4-T_3)$, 
$V(.,T_4-T_2)$, and $V(.,T_4-T_1)$, consuming four times more storage space than our method.

\subsection{Planar flight example}
\begin{figure}[H]
	\centering
	\includegraphics[width=0.43\textwidth]{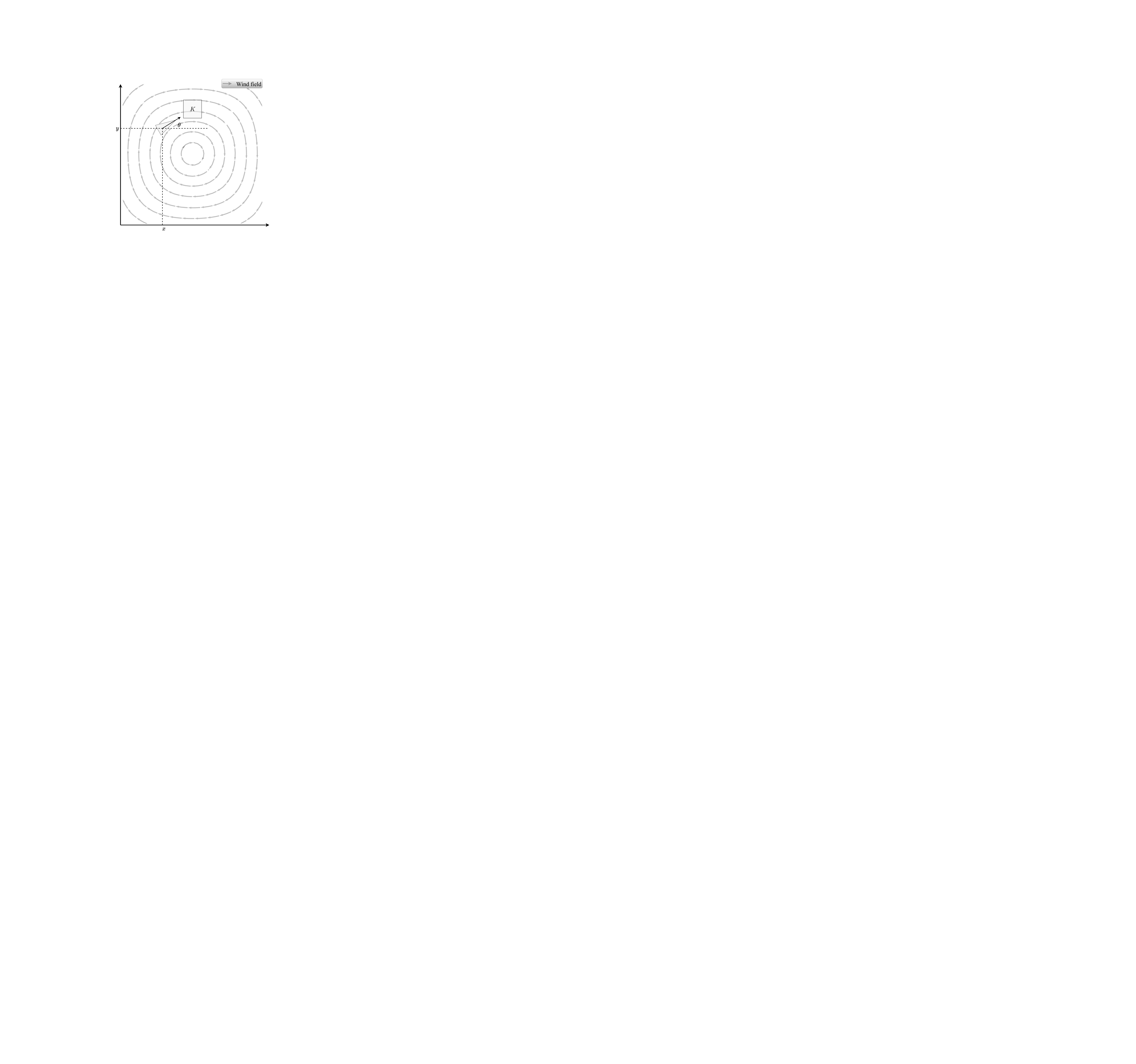}
	\caption{Visualization of the planar flight example.}
	\label{fig7}
\end{figure}

A flight vehicle moves in a plane wind field, the vehicle is 
modeled as a simple mass point with
fixed linear velocity $v=1$ and controllable heading angular velocity. 
The behavior of the vehicle in still air is described by the following equation:
\begin{align}
	\begin{split}
		&\dot{x}=v \cos \theta \\
		&\dot{y}=v \sin \theta \\
		&\dot{\theta}=u
	\end{split}
\end{align}
where $[x,y]^\mathrm{T} \in \mathbb{R}^2$ and $\theta\in [0,2\pi]$ are the position and heading angle of the vehicle respectively,
and $u\in \mathcal{U}=[-1,1]$ is the control input. The wind speed at position $[x,y]^\mathrm{T}$ is determined by the following vector field:
\begin{align}
	\left[
		\begin{array}{c}
			w_x(x,y)\\w_y(x,y)
		\end{array}
	\right]=\left[
		\begin{array}{c}
			\displaystyle{y+0.1y^3} \\ \displaystyle{-x-0.1x^3}
		\end{array}
	\right]
\end{align}
Then, the behavior of the vehicle in the wind field is:
\begin{align}
	\dot{s}=\left[
		\begin{array}{c}
			\dot{x}\\\dot{y}\\ \dot{\theta}
		\end{array}
	\right]=\left[
		\begin{array}{c}
			v \cos \theta +w_x(x,y)\\ v\sin \theta+w_y(x,y) \\ u
		\end{array}
	\right]=f(s,u)
\end{align}
where $s=[x,y,\theta]^\mathrm{T}$. The target set $K$ and task area $\Omega$ 
depend only on $x$ and $y$ and include any positions in the 
following rectangular regions in $x\ y$ plane:
\begin{align}
	&K=\{[x,y,\theta]^\mathrm{T}|x\in [-0.5,0.5],y\in [1.5,2.5]  \}\\
	&\Omega=\{[x,y,\theta]^\mathrm{T}| x\in [-4,4],y\in [-4,4]\}
\end{align}
See Fig. \ref{fig7} for a visual depiction of the problem.

The running cost is a weighted sum of the time consumption and the length of flight path per unit time, i.e.:
\begin{align}
	c(s,u)=1+\gamma \sqrt{  \dot{x}^2+  \dot{y}^2  }
\end{align}
where $\gamma$ is the weight of the length of flight path. The admissible costs are $J_1=0.75, J_2=1.5, J_3=2.25, J_4=3$. 
We consider two cases: 
\begin{itemize}
	\item [(1)] $\gamma=0$ and $\Phi(s)\equiv 0$.
	\item [(2)] $\gamma=0.1$ and $\Phi(s)=- e^{-x^2-y^2- \min( \theta,2\pi-\theta ) }$
\end{itemize}
\subsubsection{Case (1)}
In the first case, the running cost function is constant equal to 1 and the endpoint cost is constant equal to 0, 
the cost-limited reachable sets degenerate into reachable set.
The solver setups of this case are summarized in Table \ref{tb2}.
\begin{table}[H]
	\centering
	\caption{Solver settings for the first case of the planar flight example}
	\label{tb2}
	\begin{tabular}{lc}
	\toprule[1pt] 
	\textbf{Parameter} 			  			& \textbf{Setting}  \\ \toprule
	Computational domain $\Omega$         	&      $[-4,4]\times [-4,4]\times [0,2\pi]$      \\   
	Number of grid points $N_x\times N_y \times N_\theta$   &      $257\times 257 \times 257$     \\  
	$\bar{T}$               				&      $3.1$      \\    
	Number of time steps $m$            	&      $155$     \\ 
	Time step size $\Delta t$            	&      $0.02$     \\ \bottomrule[1pt] 
	\end{tabular}
\end{table}

\begin{figure}[H]
	\centering
	\subfigure[$\mathcal{R}_c(K,J_1)$]{\includegraphics[width=0.24\textwidth]{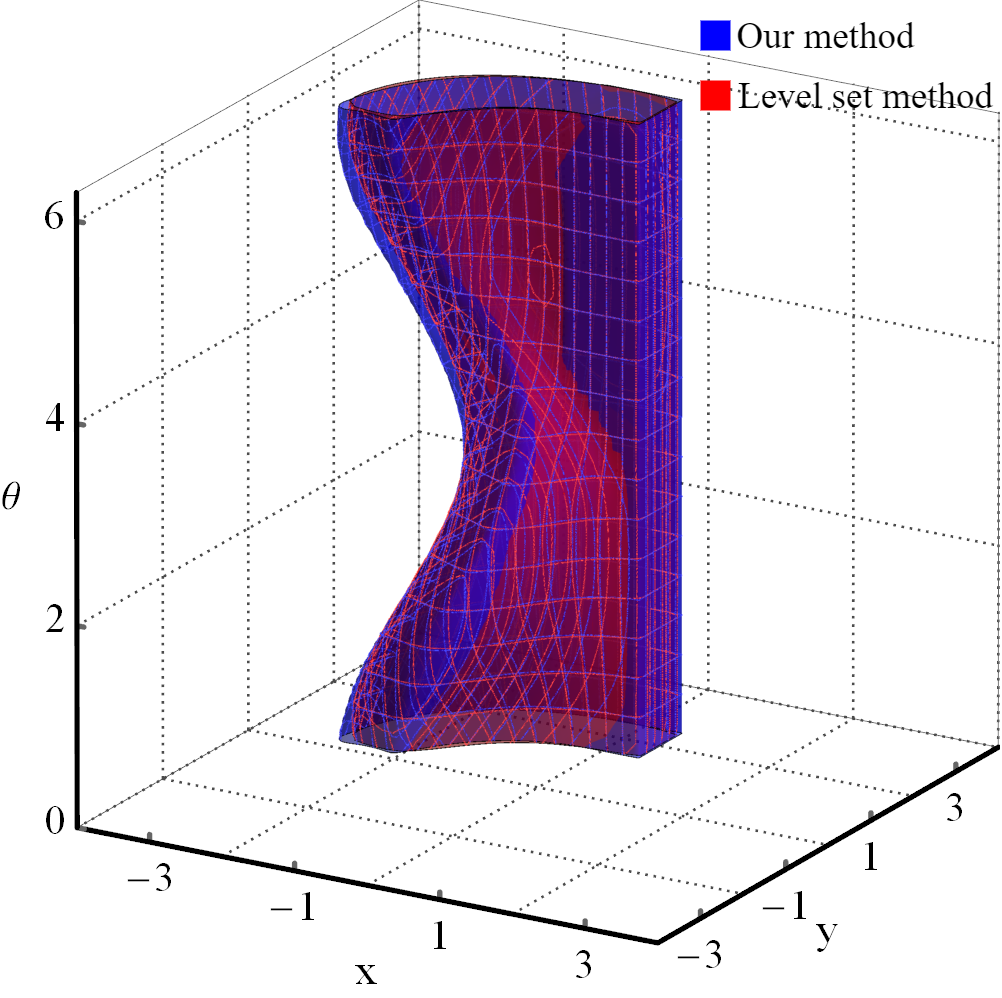}}
	\subfigure[$\mathcal{R}_c(K,J_2)$]{\includegraphics[width=0.24\textwidth]{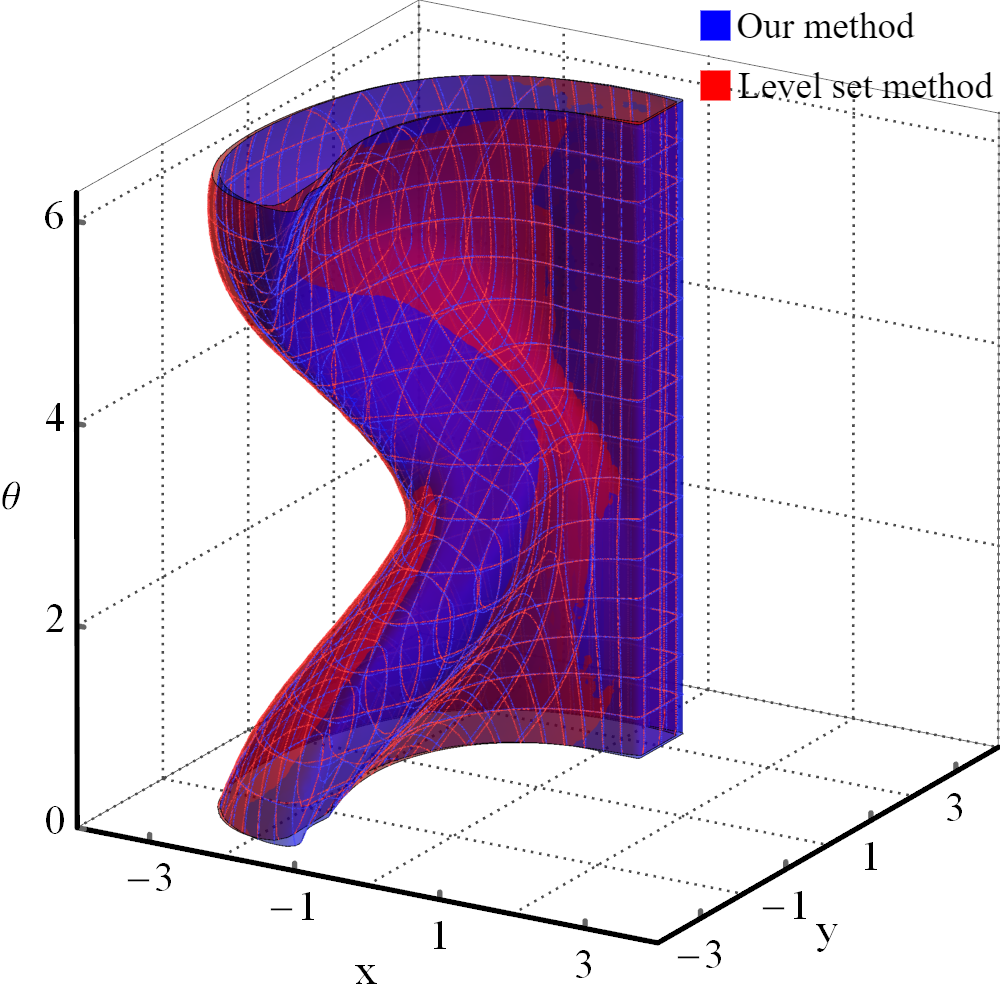}}\\
	\subfigure[$\mathcal{R}_c(K,J_3)$]{\includegraphics[width=0.24\textwidth]{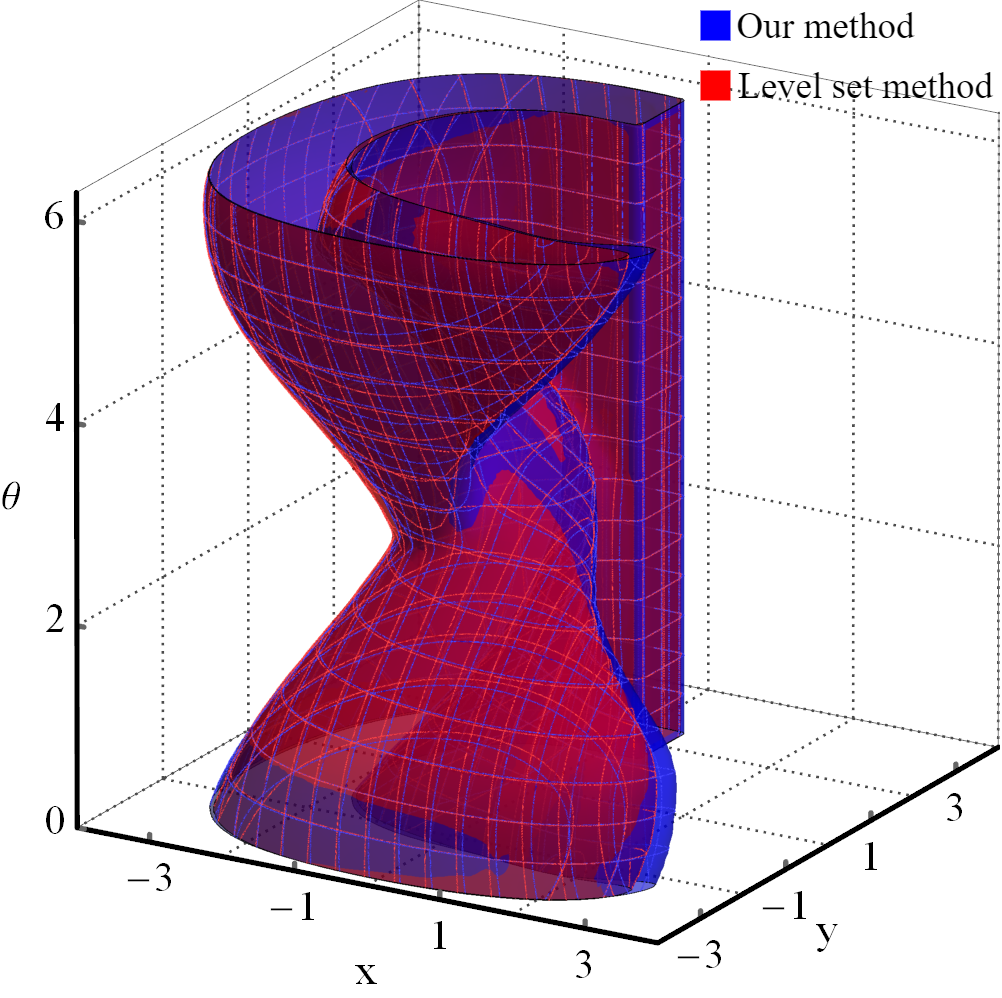}}
	\subfigure[$\mathcal{R}_c(K,J_4)$]{\includegraphics[width=0.24\textwidth]{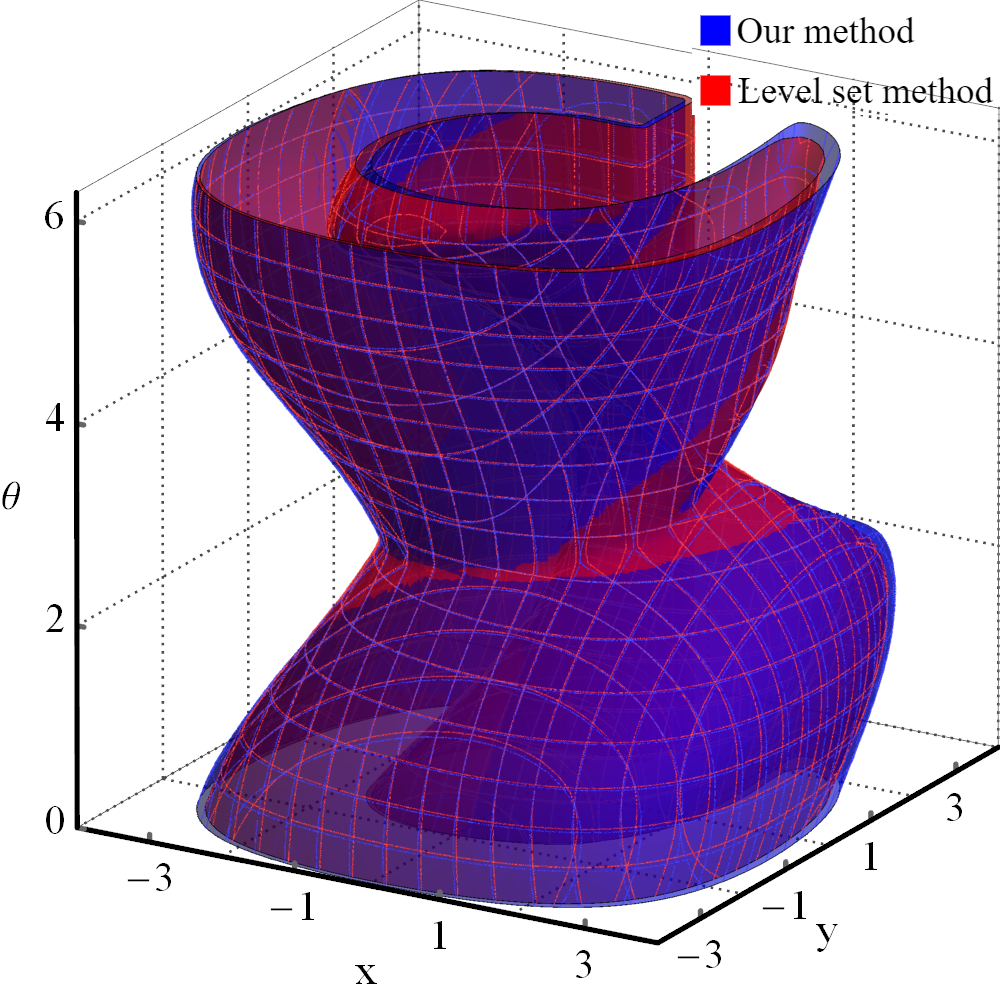}}
	\caption{Computation results of the first case of the planar flight example.}
	\label{fig8}
\end{figure}

Fig. \ref{fig8} shows the computational results of our method and the comparison with the level set method. 
As can be seen, in this example, the results of the proposed method and the level set method are also in excellent agreement.

\subsubsection{Case (2)}
In this case, the cost-limited reachable sets no longer 
degenerate to reachable sets and therefore cannot be computed using the level set method,
but can still be computed using the proposed method. 

\begin{figure}[H]
	\centering
	\subfigure[$\mathcal{R}_c(K,J_1)$]{\includegraphics[width=0.24\textwidth]{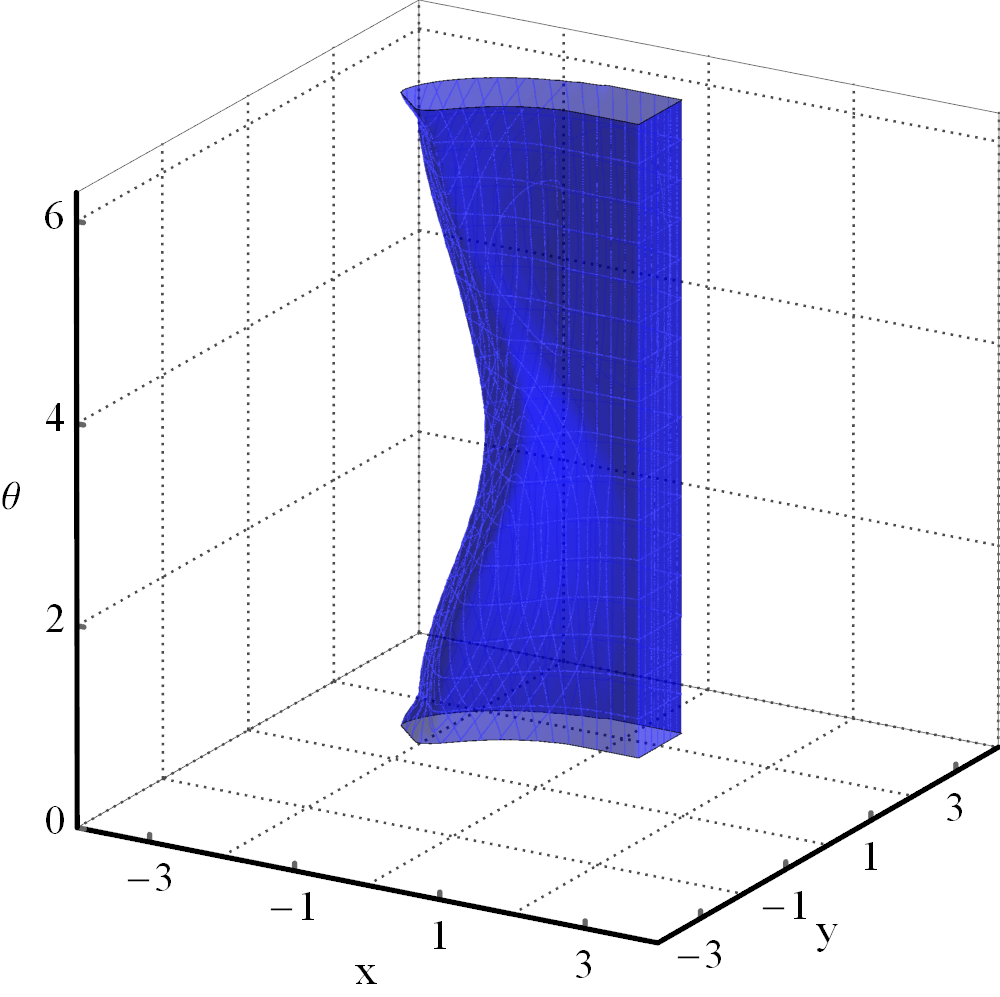}}
	\subfigure[$\mathcal{R}_c(K,J_2)$]{\includegraphics[width=0.24\textwidth]{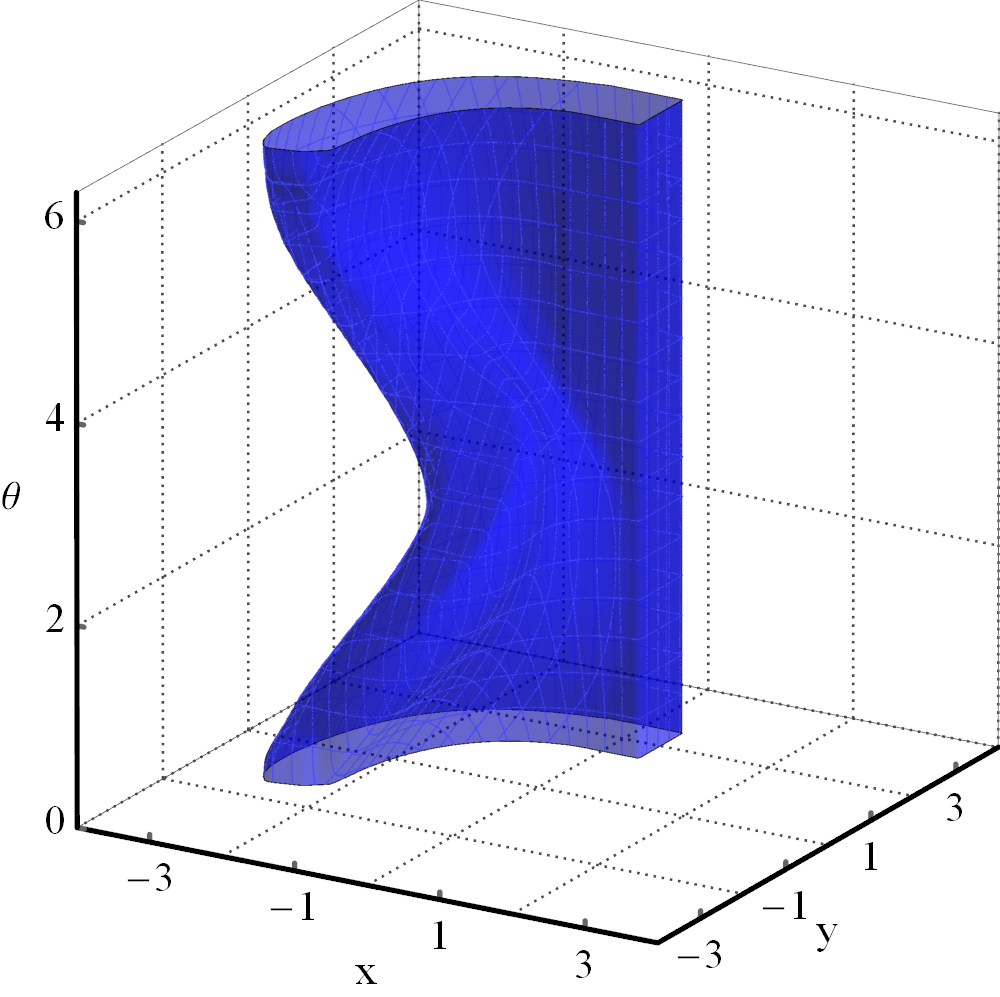}}\\
	\subfigure[$\mathcal{R}_c(K,J_3)$]{\includegraphics[width=0.24\textwidth]{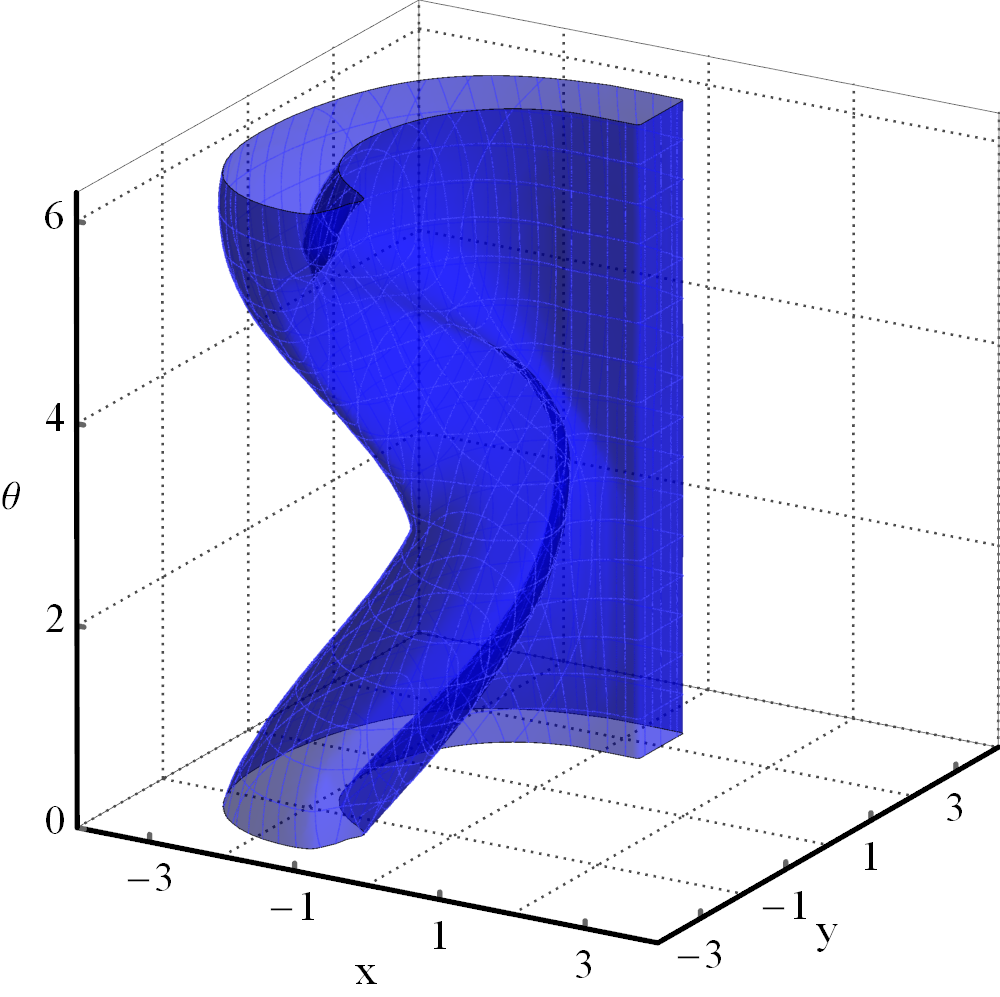}}
	\subfigure[$\mathcal{R}_c(K,J_4)$]{\includegraphics[width=0.24\textwidth]{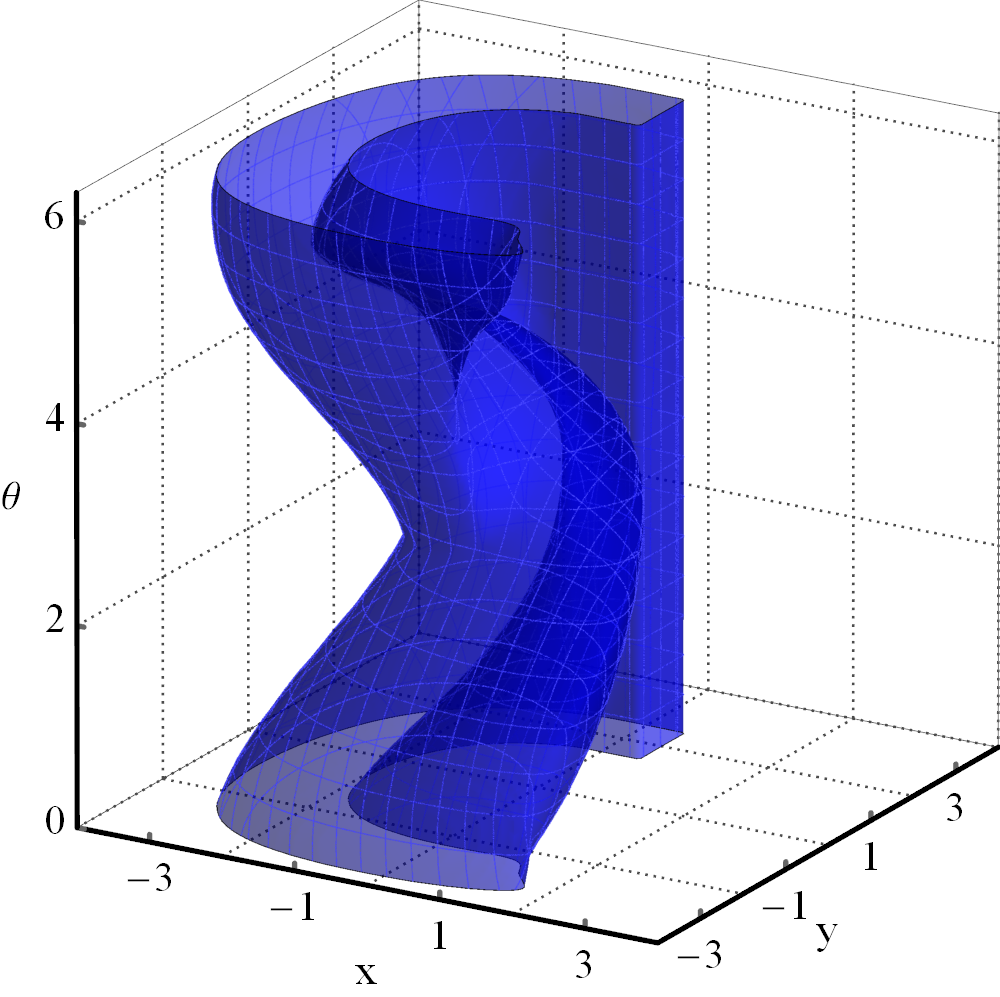}}
	\caption{Computation results of the second case of the planar flight example.}
	\label{fig9}
\end{figure}

In this case, 
$\lambda=\displaystyle{ \min_{s,u}c(s,u)=1 }$, $\Lambda = \displaystyle{\min_{s} \Phi(s)=-1} $.
According to line 3 of Algorithm \ref{alg1}, $\bar{T}$ should satisfy the following inequalities:
\begin{align}
	\lambda\bar{T}+\Lambda > J_4 \Longrightarrow \bar{T}>4
\end{align}
Therefore, $\bar{T}$ is set to 4.1. 
The solver setups of this case are listed in Table \ref{tb3}.
\begin{table}[H]
	\centering
	\caption{Solver settings for the second case of the planar flight example}
	\label{tb3}
	\begin{tabular}{lc}
	\toprule[1pt] 
	\textbf{Parameter} 			  			& \textbf{Setting}  \\ \toprule
	Computational domain $\Omega$         	&      $[-4,4]\times [-4,4]\times [0,2\pi]$      \\   
	Number of grid points $N_x\times N_y \times N_\theta$   &      $257\times 257 \times 257$     \\  
	$\bar{T}$               				&      $4.1$      \\    
	Number of time steps $m$            	&      $205$     \\ 
	Time step size $\Delta t$            	&      $0.02$     \\ \bottomrule[1pt] 
	\end{tabular}
\end{table}
The computation results are shown in Fig. \ref{fig9}.

\begin{figure}[H]
	\centering
	\subfigure[Slice of $\mathcal{R}_c(K,J_1)$]{\includegraphics[width=0.24\textwidth]{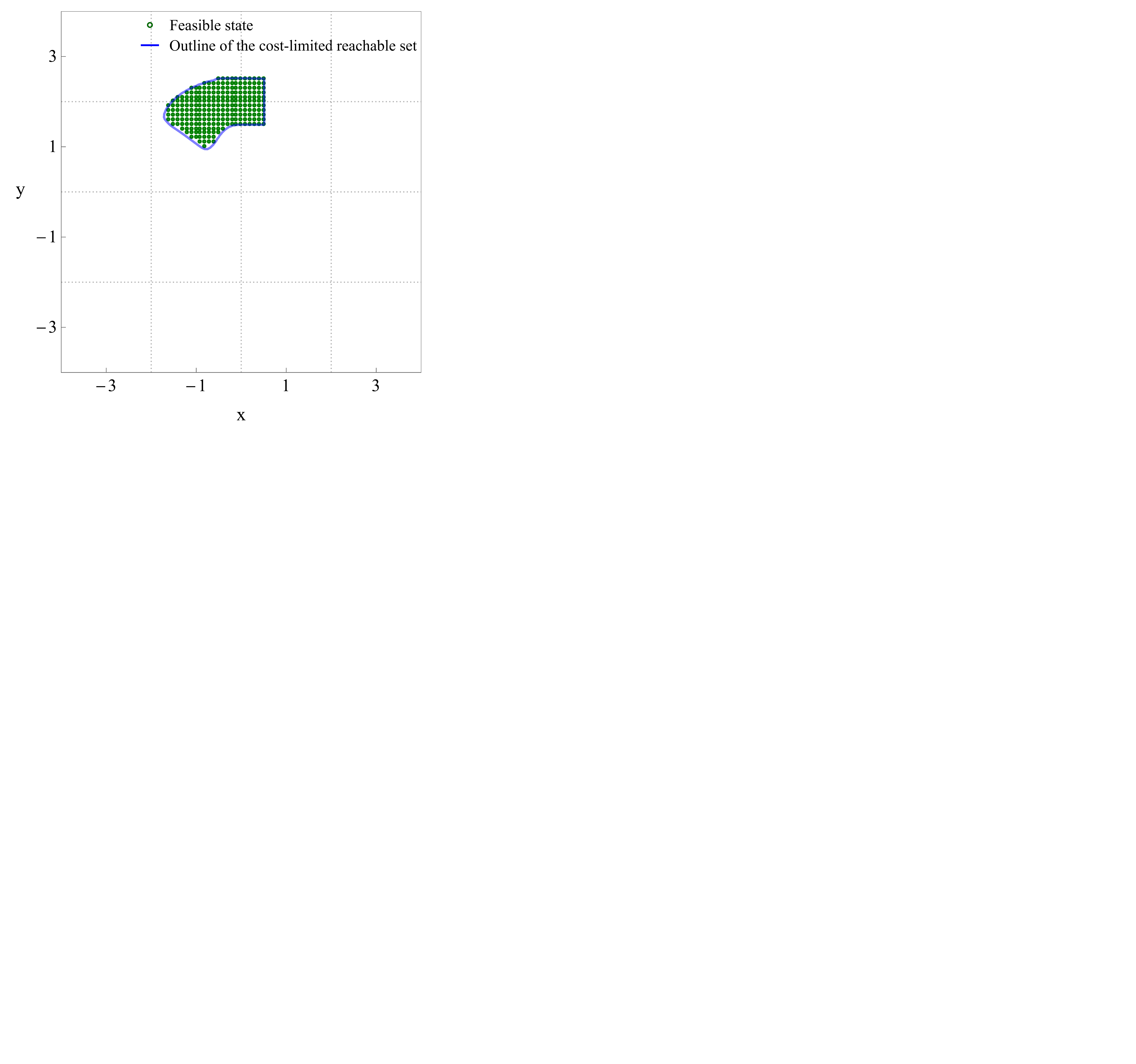}}
	\subfigure[Slice of $\mathcal{R}_c(K,J_2)$]{\includegraphics[width=0.24\textwidth]{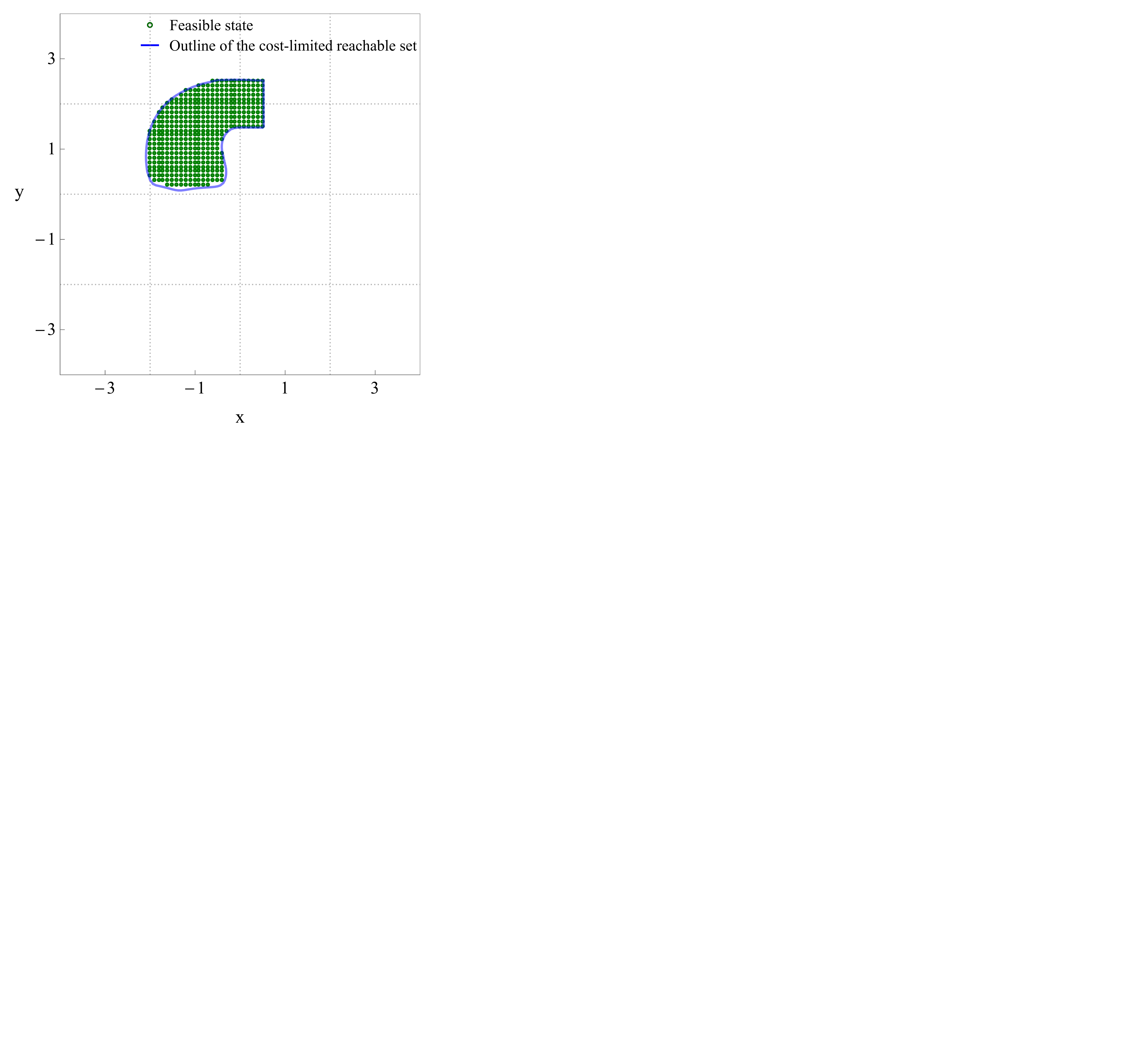}}\\
	\subfigure[Slice of $\mathcal{R}_c(K,J_3)$]{\includegraphics[width=0.24\textwidth]{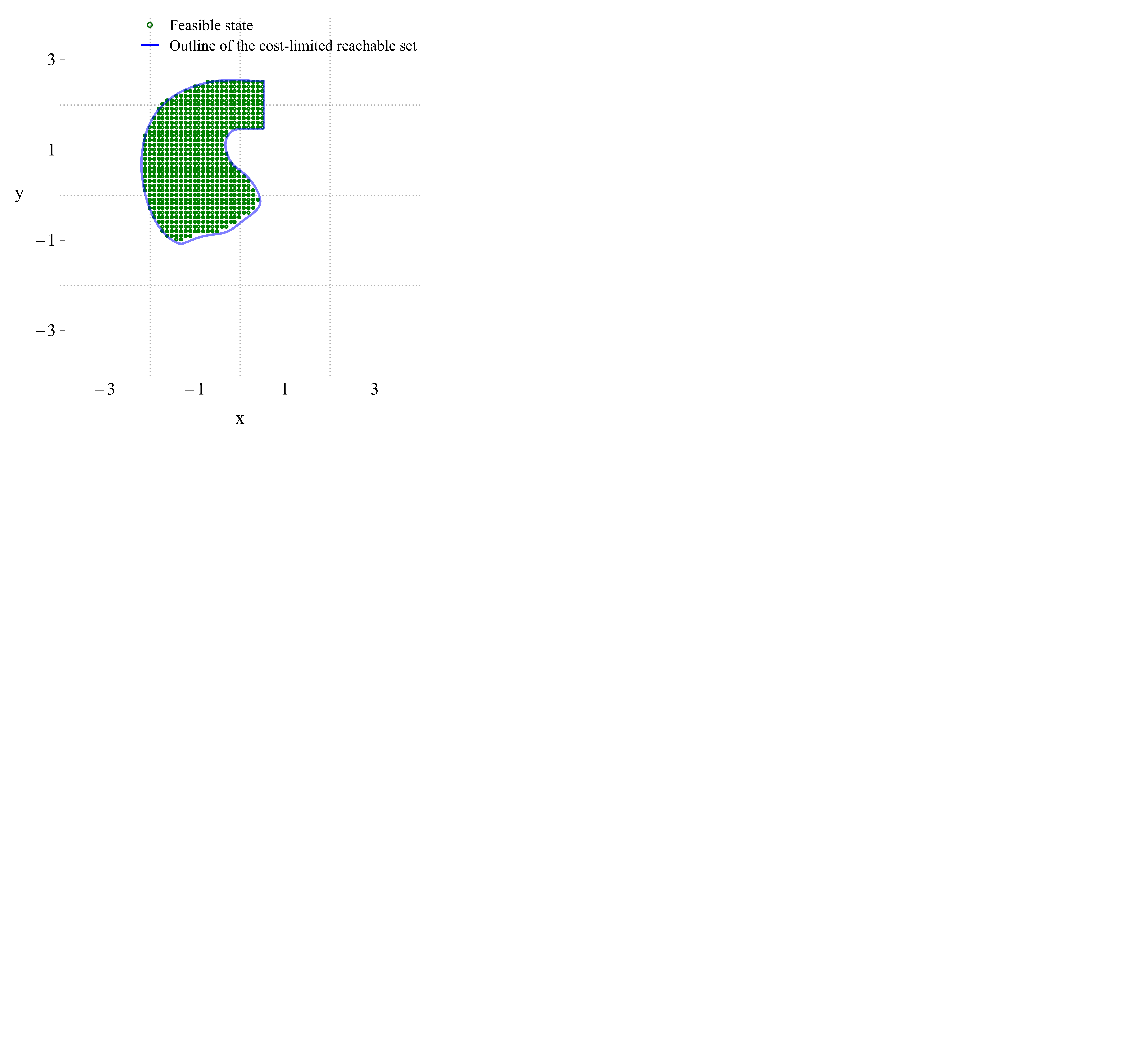}}
	\subfigure[Slice of $\mathcal{R}_c(K,J_4)$]{\includegraphics[width=0.24\textwidth]{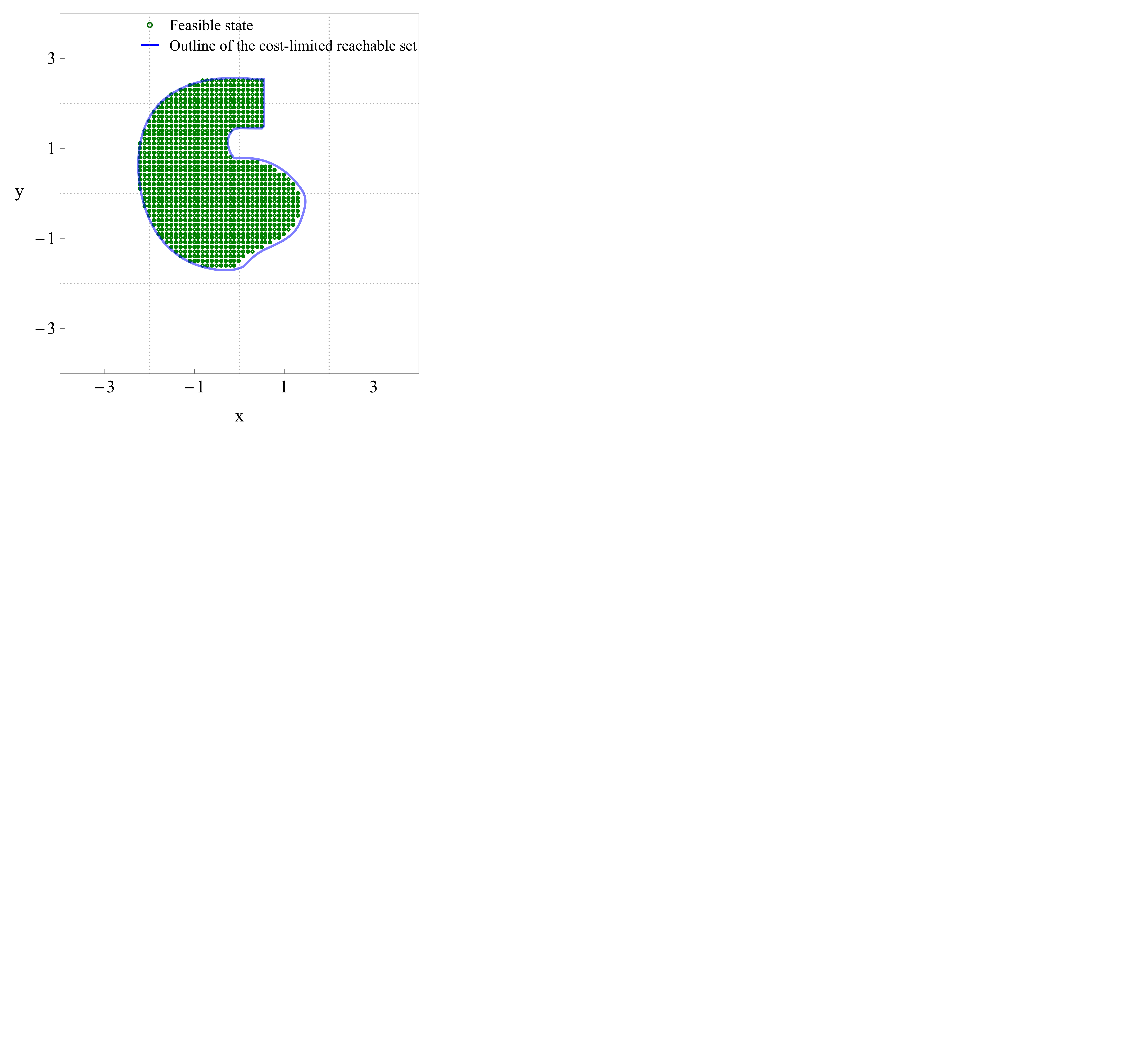}}
	\caption{Verification results.}
	\label{fig10}
\end{figure}

In order to verify the correctness of the results and the validity of the control law 
in Eq. (\ref{cl3}), 
we test some states in slice $ \theta=\pi $ of the state space
to determine whether the trajectories initialized from these states can reach the target set before 
the performance index increasing to the given admissible costs under this control law.
The verification results are shown in Fig. \ref{fig10}.
It can be seen that the outlines of the cost-limited reachable sets and the borders of the areas marked
by the green circles almost coincide, which shows the accuracy of the computation of the 
cost-limited reachable sets and the validity of the control law in Eq. (\ref{cl3}).

\section{Conclusions}
This paper proposes a new method for computing reachable sets. In the proposed method, 
the reachable sets of different time horizons are represented by different non-zero level sets of 
a HJB equation with a running cost function. 
This approach significantly reduces the storage space consumption for saving reachable sets and designing control laws.

In addition to being able to solve the classical reachability problems, the proposed method can also solve more generalized 
reachability problems by setting different operating cost functions and different terminal conditions for the HJB equation.
The reachable sets in such problems are referred to in this paper as cost-limited reachable sets

In order to overcome the discontinuity of the solution of the HJB equation, 
the current research adopts a method based on recursion and grid interpolation for solving the HJB equation.
The paper concludes with some examples to illustrate the effectiveness and generality of the proposed method.

However, the proposed method has some potential for improvement. 
The main drawback of the proposed method, and also the main drawback of the level set method, lies in
the exponential growth of memory and computational cost as the system dimension increases.
Some approaches have been proposed to mitigate these costs,
such as splitting the original high-dimensional system into multiple 
low-dimensional subsystems based on the dependencies between the system states \cite{a6,i5} or the time-scale principle \cite{a18,i11}. 
These approaches will be considered in our future works.




\section*{Acknowledgements}
The authors would like to thank
the anonymous reviewers, associate editor, and editor for
their valuable and constructive comments and suggestions.

\bibliographystyle{Bibliography/IEEEtranTIE}
\bibliography{mybib2}

\end{document}